\documentclass[aps,prd,reprint,superscriptaddress]{revtex4-2}
\usepackage{graphicx}
\usepackage{amsmath,amssymb,amsfonts}
\usepackage{bm}
\usepackage{physics}
\usepackage{xcolor}
\usepackage{mathtools}
\usepackage{float}
\usepackage{hyperref}

\begin{document}

\title{First broadband search for fixed-polarization dark photon dark matter with GigaBREAD}

\author{Jialin Yu}
\email[Corresponding author: ]{jyu53@hawk.illinoistech.edu}
\affiliation{Department of Physics, Illinois Institute of Technology,
Chicago, Illinois 60616, USA}
\affiliation{Fermi National Accelerator Laboratory,
Batavia, Illinois 60510, USA}
\author{Stefan Knirck}
\email{knirck@fas.harvard.edu}
\affiliation{Laboratory for Particle Physics and Cosmology,
Department of Physics, Harvard University,
Cambridge, Massachusetts 02138, USA}

\author{Gabe Hoshino}
\altaffiliation{Now at: University of Hamburg}
\affiliation{Department of Physics, University of Chicago,
Chicago, Illinois 60637, USA}
\affiliation{Enrico Fermi Institute, University of Chicago,
Chicago, Illinois 60637, USA}
\affiliation{Kavli Institute for Cosmological Physics,
University of Chicago, Chicago, Illinois 60637, USA}

\author{Rakshya Khatiwada}\affiliation{Department of Physics, Illinois Institute of Technology,
Chicago, Illinois 60616, USA}
\affiliation{Fermi National Accelerator Laboratory,
Batavia, Illinois 60510, USA}

\author{Alex Lapuente}
\altaffiliation{Now at: Physics Department, Boston University, Boston, MA 02215, USA.}
\affiliation{Department of Physics, University of Chicago,
Chicago, Illinois 60637, USA}

\author{David W. Miller}
\affiliation{Department of Physics, University of Chicago,
Chicago, Illinois 60637, USA}
\affiliation{Enrico Fermi Institute, University of Chicago,
Chicago, Illinois 60637, USA}
\affiliation{Kavli Institute for Cosmological Physics,
University of Chicago, Chicago, Illinois 60637, USA}

\author{Andrew Sonnenschein}
\affiliation{Fermi National Accelerator Laboratory,
Batavia, Illinois 60510, USA}
\affiliation{Enrico Fermi Institute, University of Chicago,
Chicago, Illinois 60637, USA}

\date{\today}

\begin{abstract}
We present the first search for fixed-polarization dark photon (DP) dark matter in the 44--52~$\mu\mathrm{eV}$ mass range, corresponding to 10.7--12.5~GHz. GigaBREAD is a room-temperature broadband dish antenna experiment in which DP convert to ordinary photons on a cylindrical metallic emission surface and are focused onto a radio-frequency receiver. Previous searches, including GigaBREAD and several other dish antenna searches, have assumed randomly polarized DP dark matter. However, vector dark matter produced by a misalignment-like mechanism may retain an approximately fixed polarization, leading to sidereal modulation of the detector response as the Earth rotates. We reanalyze the GigaBREAD data using a polarization-matched likelihood that accounts for the time-dependent projection of a fixed DP polarization onto the detector axis. We find no evidence for a fixed-polarization DP signal and set upper limits on the kinetic mixing parameter at the level of $\chi\sim10^{-12}$. These limits are about $20\%$ better than a naive limit obtained by rescaling the previous GigaBREAD limit under the fixed-polarization assumption.
\end{abstract}

\maketitle

\section{Introduction}

The existence of dark matter is supported by a wide range of astrophysical and cosmological observations, yet its composition remains unknown \cite{RubinFord1970,Tyson1998,Tegmark2004}. One broadly studied class of candidates consists of light, weakly interacting bosonic dark matter, such as axions and dark photons (DP).
Similar to axions, DPs constitute another class of light bosonic dark matter candidates, typically described as gauge bosons of an additional U(1) symmetry beyond the Standard Model. DPs can couple feebly to ordinary photons through kinetic mixing, producing photons with frequencies defined by DP mass. Since this coupling does not rely on an applied magnetic field, many haloscope experiments originally designed for axion searches could also probe DPs. 

Ultralight dark-photon dark matter can be described by a classical massive vector field with two transverse polarization modes and one longitudinal mode. The polarization distribution of relic DP dark matter depends on its production mechanism. In particular, misalignment-like production can yield a polarization direction that is coherent within a cosmological horizon. Although structure formation may modify the primordial polarization distribution, a direction that remains effectively fixed over laboratory scales and observation times is a plausible scenario. As the Earth rotates, the detector projection onto this fixed direction varies with sidereal time~\cite{Caputo2021}.

The sensitivity of a DP search depends on the angle $\vartheta$ between the DP polarization and the detector's sensitivity axis. In particular, 
most detection experiments are sensitive to a single linear polarization direction in the detector frame, so the detected signal can be written as:

\begin{equation}
P_{\mathrm{det}}(t)=P_0\cos^2\vartheta(t)=P_0\bigl(\hat{\mathbf{X}}\!\cdot\!\hat{\mathbf{D}}(t)\bigr)^2,
\end{equation}
where $\hat{\mathbf{X}}$ is the polarization direction and $\hat{\mathbf{D}}(t)$ is
the detector sensitivity axis. 

Previous DP searches, for example GigaBREAD \cite{GigaBREAD2024_PRL}, QUALIPHIDE \cite{QUALIPHIDE2023}, and SHUKET \cite{SHUKET2019}, have typically assumed a randomly oriented DP polarization. In this case, the relevant projection factor averages to $\langle \cos^2\vartheta \rangle = 1/3$ over time, since a random polarization corresponds to equal probability along three orthogonal directions.
This average is then used as a constant suppression factor in exclusion limit calculations. However, this assumption has been challenged, as it may not accurately reflect the true nature of the DP signal, potentially introducing biases in exclusion limit estimates \cite{Caputo2021}.

\begin{figure}[tb]
  \centering
  \includegraphics[width=0.5\textwidth]{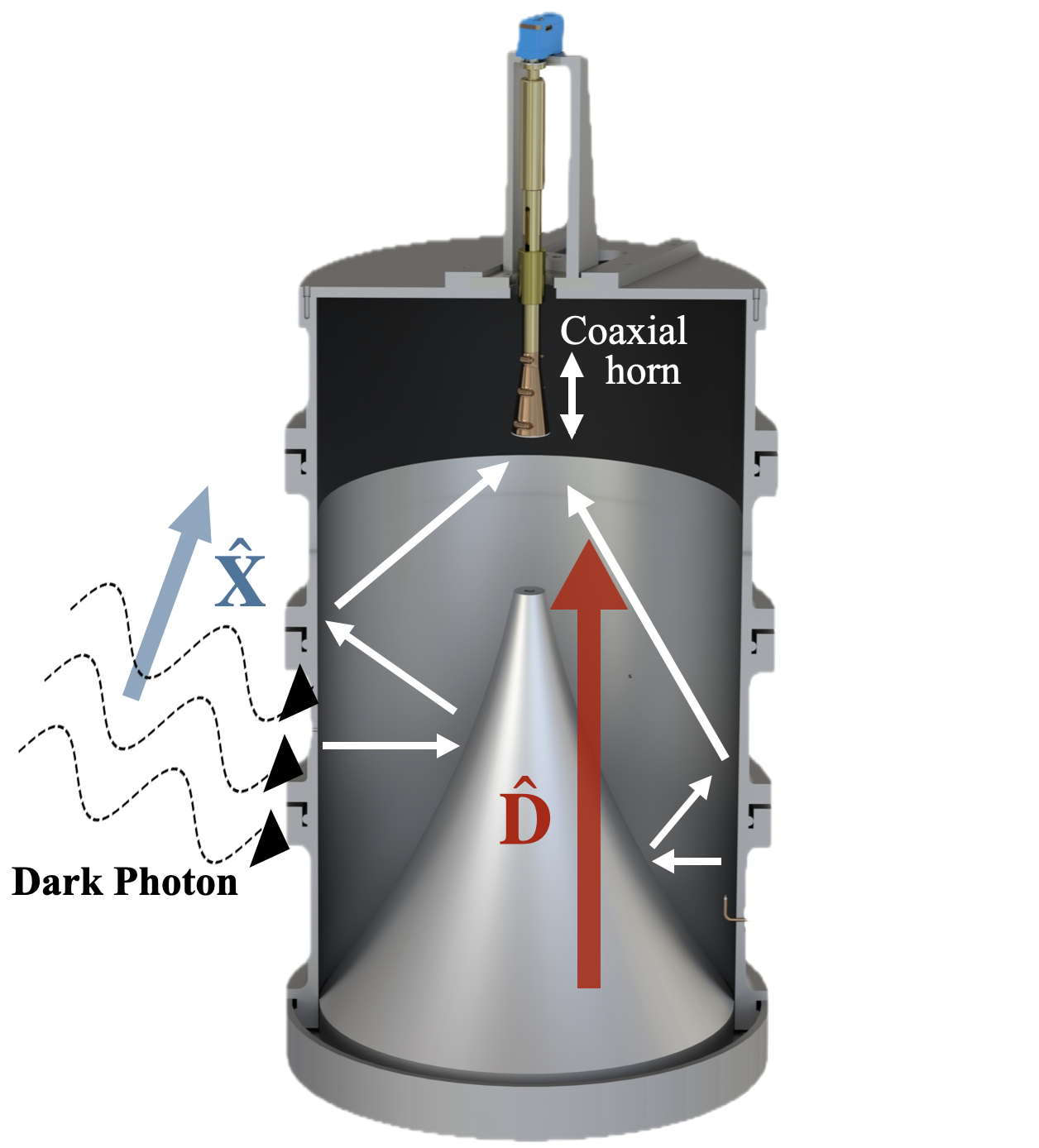}
  \caption{GigaBREAD reflector setup. When dark photons pass through the vertical metallic cylinder, electric fields parallel to its surface induce electron oscillations, but the coaxial horn antenna exclusively couples to the vertically polarized emissions. This geometric constraint strictly defines the experiment's vertical sensitivity axis (red arrow). The oscillating electrons then emit ordinary photons perpendicularly (horizontally) toward the central cone, which reflects them upward into the coaxial horn for detection}
  \label{fig:BREAD}
\end{figure}

Cavity haloscopes rely on the resonant enhancement of specific cavity modes to increase the probability of dark matter converting into detectable photons~\cite{ADMX2021,DFSZ2023,HAYSTAC2026,HAYSTAC2023,CAPP2024,PhysRevD.110.043022}. Dish antennas detect electromagnetic waves emitted perpendicularly from a metallic surface, which occur when a dark photon's electric field causes electrons in the dish to oscillate~\cite{Suzuki_2015,SHUKET2019,Tomita_2020,QUALIPHIDE2023,DOSUE2023,CMWR2023,BRASS-p2023}. Unlike cavities, which scan over a frequency range, broadband dish antenna experiments such as BREAD take full-frequency spectrum measurements in each acquisition. As shown in Figure~\ref{fig:BREAD}, GigaBREAD has a vertical metallic cylinder and a central reflective cone to focus dark photon-induced emissions into a coaxial horn, establishing a strictly vertical sensitivity axis. The experiment can operate for extended periods, up to several months, resulting in tens of thousands of data acquisitions. Each acquisition yields a time-stamped power spectrum. A fixed DP polarization can therefore produce a time-dependent signal through the projection factor $\cos^2\vartheta(t)$.

In this work, we apply the time-dependent, integrated-likelihood analysis to two GigaBREAD data sets acquired in the Chicago area: a zenith--aligned run on the University of Chicago campus (GigaBREAD2023), and a horizontal-axis run at Argonne National Laboratory (GigaBREAD2024). The experimental hardware and calibration procedures are the same as in publications~\cite{GigaBREAD2024_PRL, ArgonneAxion2025}; here we summarize only the geometry, site coordinates, and data products used by the statistical framework (shown in Table~\ref{tab:daq_runs}).

\begin{table*}[htbp]
\centering
\renewcommand{\arraystretch}{1.3}
\begin{tabular}{|l|c|c|l|}
\hline
\textbf{Run Name} & \textbf{Science Data} & \textbf{Position (Lat/Long)} & \textbf{Alignment Direction} \\
\hline
GigaBREAD2023 & 24 days& $41.79^\circ$\,N, $87.60^\circ$\,W & Vertical, $\hat{\mathbf{Z}}$ \\
\hline
GigaBREAD2024 &27 days & $41.71^\circ$\,N, $87.98^\circ$\,W &  Horizontal: $\hat{\mathbf{V}}$, $\phi_d = -55^\circ$\\
\hline
\end{tabular}
\caption{Summary of two GigaBREAD runs}
\label{tab:daq_runs}
\end{table*}

During data taking, the power spectra were recorded continuously with timestamps $t_k$. The UChicago and ANL runs comprise 12,970 and 12,481 spectra, corresponding to total observing times of 24 days and 27 days. Each spectrum consists of $230{,}392$ uniformly spaced frequency bins $f_j$ with a resolution of $\Delta f = 7.8$\,kHz, covering the 10.7–12.5\,GHz range~\cite{GigaBREAD2024_PRL, ArgonneAxion2025}. Here $k$ indexes the recorded spectra and $j$ indexes the frequency bins.

Each recorded power spectrum $P_{{\rm SA}, kj} \equiv P_{\rm SA}(t_k, f_j)$ was smoothed across the frequency domain using a fourth-order Savitzky-Golay filter to derive the instrumental baseline $P_{{\rm bl}, kj}$~\cite{HAYSTAC_procedure}. The expected standard deviation of the residual noise is given by the radiometer equation, $\sigma_{kj} = P_{{\rm bl}, kj}/\sqrt{N_{\text{av}, kj}}$, where $N_{\text{av}}$ is the number of averages. The residual power is then extracted by subtracting the baseline from the raw spectrum: $y_{kj} = P_{{\rm SA}, kj} - P_{{\rm bl}, kj}$.

This work presents an analysis of fixed-polarization DP signals and applies it to the GigaBREAD data. The resulting fixed-polarization limit is comparable to the previous random-polarization limit. The paper is structured as follows: Section~\ref{sec:model} summarizes the expected DP signal model, including conversion power, polarization modulation, and the frequency-bin response.
Section~\ref{sec:analysis_likelihood} presents the statistical framework and test statistics.
Section~\ref{sec:search_results} applies the method to the GigaBREAD data and presents results, with supplementary derivations and validation details given in the appendices.

\section{Expected Dark-Photon Signal Model}
\label{sec:model}

 In a GigaBREAD experimental setup, the continuous power signal is digitized and Fourier-transformed by the Data Acquisition (DAQ) system, generating a sequence of discrete power spectra. We model the expected discrete signal $\mathcal{S}_{kj}$ in the $j$-th frequency bin of the $k$-th time bin as follows:

\begin{equation}
\mathcal{S}_{kj} = P_{0}(\chi, m_{\mathrm{DP}})\,\eta_{kj}\,\mathcal{C}_j(m_{\mathrm{DP}})\, \cos^2\!\vartheta(t_k), \label{eq:Skj}
\end{equation}

where the physical and instrumental factors entering this signal model are grouped as follows:

\begin{enumerate}
    \item \textbf{Conversion Power and Efficiency:} deriving the overall signal amplitude $P_{0}\eta$ from the kinetic-mixing mechanism and experimental parameters;
    \item \textbf{Polarization Modulation:} describing the time-dependent factor $\cos^2\!\vartheta(t)$ arising from the relative geometry between the DP polarization and the detector orientation;
    \item \textbf{Frequency-Bin Response:} defining $\mathcal{C}_j(f_0)$ from the intrinsic dark-matter lineshape and the cross-correlated FFT-bin response.
\end{enumerate}

\subsection{Conversion Power and Efficiency}
\label{subsec:amp}

For a reflective surface much larger than the photon wavelength and with negligible roughness, dark-photon conversion is non-resonant and broadband. The reference conversion power \(P_0\) has no explicit mass dependence, although the detector efficiency can vary with frequency.
We define $P_0$ as the reference conversion power for unit polarization projection, corresponding to perfect alignment between the DP polarization and the detector sensitivity axis:~\cite{Caputo2021, GigaBREAD2024_PRL}
\begin{equation}
\begin{aligned}
P_0 &= 1.8\times10^{-20}\,{\rm W}  \left(\frac{A}{0.5\,{\rm m^2}}\right) \left(\frac{\chi}{10^{-12}}\right)^2 \\ 
&\quad\times \left(\frac{\rho_{\rm DM}}{0.45\,{\rm GeV\,cm^{-3}}}\right) 
\end{aligned}
\label{eq:Pconv}
\end{equation}
Here, $A$ is the effective emission area of the dish, $\chi$ is the kinetic-mixing strength between photons and DPs, and $\rho_{\rm DM}$ is the local dark-matter density (taken here as $0.45\,{\rm GeV\,cm^{-3}}$~\cite{Read2014, PDG2024}). This converted power is then coupled into the horn antenna, yielding a detectable signal $\eta P_0$, where $\eta$ is the detection efficiency that varies with the position of the horn.

\subsection{Polarization Modulation}
\label{subsec:mod}

The polarization of a DP field can point in any direction, specified by its latitude $\theta_X$ and azimuth $\phi_X$ in the geocentric equatorial frame. To evaluate the polarization factor $\cos^2\!\vartheta(\theta_X,\phi_X,t)$, we express this direction as a unit vector $\hat{\mathbf{X}}$ in the basis $(\hat{x}_e,\hat{y}_e,\hat{z}_e)$. 
We write the fixed polarization of the DM field as ~\cite{Caputo2021}:
\begin{equation}
\hat{\mathbf{X}} =
\begin{pmatrix}
\cos \theta_X \cos \phi_X \\
\cos \theta_X \sin \phi_X \\
\sin \theta_X
\end{pmatrix},
\end{equation}

Most DP searches using directional detectors are sensitive primarily to the field component induced along a single sensitivity axis. 
For example, the GigaBREAD2023  run was zenith-aligned, and its sensitivity direction can be expressed as $\hat{\mathbf{Z}}(t)$~\cite{GigaBREAD2024_PRL}. 
If the detector is instead oriented horizontally (parallel to the ground), as in the GigaBREAD2024 run, the sensitivity direction is described by $\hat{\mathbf{V}}(\phi_d,t)$, where $\phi_d$ denotes the detector’s azimuthal orientation~\cite{ArgonneAxion2025}.

\begin{figure*}[t]
\centering
\includegraphics[height=0.35\textwidth]{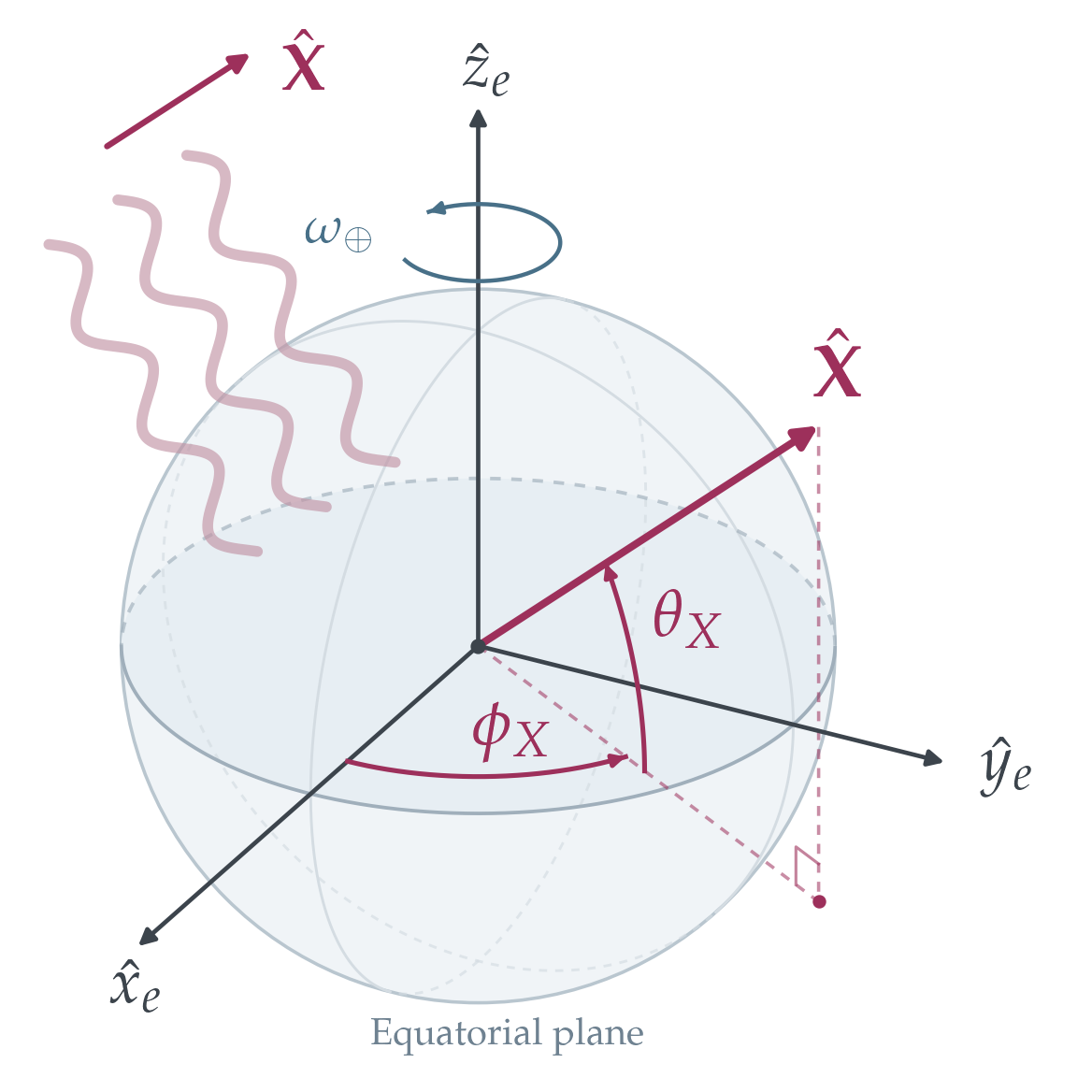}
\includegraphics[height=0.35\textwidth]{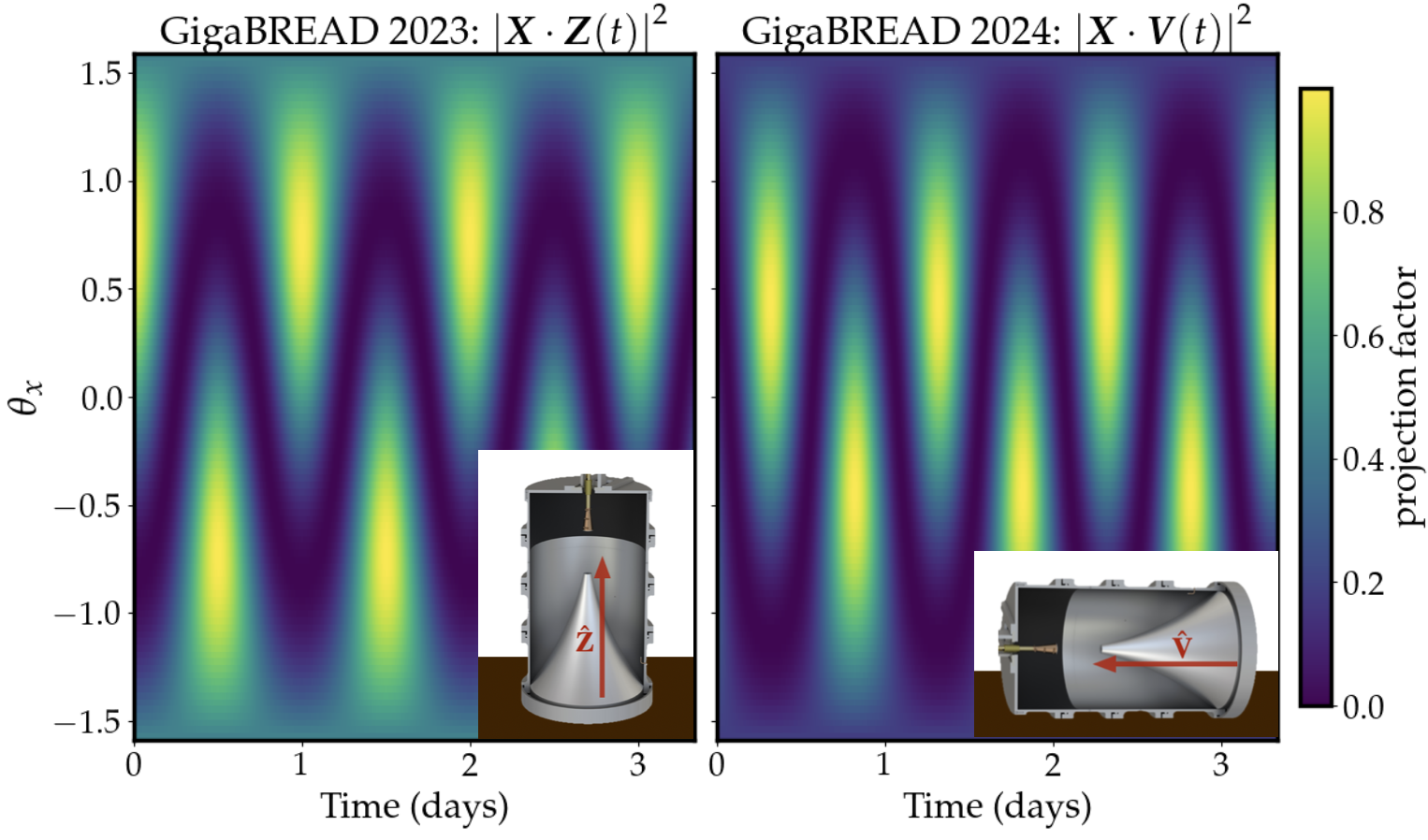}
    \caption{
    \textbf{Left:} Coordinate systems used in this analysis. 
    The geocentric equatorial frame is shown, with the DP polarization vector $\hat{\mathbf{X}}(\theta_X,\phi_X)$ fixed in the galactic frame. 
    The $\hat{x}_e$ axis is conventionally aligned toward the vernal equinox, and $\hat{y}_e$ points $90^\circ$ east in right ascension. 
    The Earth rotates counterclockwise in the $\hat{x}_e$–$\hat{y}_e$ plane. 
    The visualization is adapted from Ref.~\cite{Caputo2021}, note that the polar angle $\theta_X$ is defined here as the latitude from the equatorial plane.
    \textbf{Middle:} Example of sidereal modulation $|\hat{\mathbf{X}}\cdot\hat{\mathbf{Z}}(t)|^2$ in the zenith-aligned (vertical) configuration (GigaBREAD2023 Uchicago run).
    \textbf{Right:} Modulation pattern $|\hat{\mathbf{X}}\cdot\hat{\mathbf{V}}(\phi_d,t)|^2$ for the horizontal configuration (GigaBREAD2024 ANL run).
     }
    \label{fig:earth and modualtions}
\end{figure*}

For the vertical run the experiment is sensitive to polarization along:
\begin{equation}
\hat{\mathbf{Z}}(t) =
\begin{pmatrix}
\cos \lambda \cos (\omega_\oplus t) \\
\cos \lambda \sin (\omega_\oplus t) \\
\sin \lambda
\end{pmatrix},
\label{eq:XZ_vecs}
\end{equation}
and in horizontally run, which is perpendicular to $\hat{\mathbf{Z}}(t)$: 
\begin{equation}
\hat{\mathbf{V}}(\phi_d, t) =
\begin{pmatrix}
\cos \phi_d \sin \lambda \cos (\omega_\oplus t) + \sin \phi_d \sin (\omega_\oplus t) \\
\cos \phi_d \sin \lambda \sin (\omega_\oplus t) - \sin \phi_d \cos (\omega_\oplus t) \\
-\cos \phi_d \cos \lambda
\end{pmatrix},
\label{eq:V_vec}
\end{equation}
where $\lambda$ is the latitude of the experimental site and $\omega_\oplus$ is the Earth's sidereal rotational frequency. 
The parameter $\theta_X$ represents the latitude of the DP polarization vector in Earth coordinates. 
It measures the angular distance between $\hat{\mathbf{X}}$ and the Earth's equatorial plane, with $\theta_X = 0^\circ$ at the equator and $\theta_X = \pm90^\circ$ at the poles, as illustrated in Fig.~\ref{fig:earth and modualtions}.

The instantaneous polarization factor can then be written as
\begin{equation}
\cos^2\!\vartheta(\theta_X,\phi_X,t) =
\begin{cases}
|\hat{\mathbf{X}}\cdot\hat{\mathbf{Z}}(t)|^2,  &\text{zenith-aligned },\\[4pt]
|\hat{\mathbf{X}}\cdot\hat{\mathbf{V}}(\phi_d,t)|^2,  &\text{horizontal}
\end{cases}
\label{eq:cos2_def}
\end{equation}

We can expand $|\hat{\mathbf{X}}\!\cdot\!\hat{\mathbf{Z}}(t)|^2$ as \[
|\hat{\mathbf{X}}\!\cdot\!\hat{\mathbf{Z}}(t)|^2 = C_{\rm const} + C_{\rm daily}(t) + C_{\rm 2-daily}(t)
\]
revealing three distinct time dependencies ~\cite{Caputo2021}:
\begin{itemize}
    \item \textbf{Constant term (time-independent):}
    \[
    C_{\text{const}} = \tfrac{1}{2}\cos^2\!\lambda\,\cos^2\!\theta_X + \sin^2\!\lambda\,\sin^2\!\theta_X.
    \]
    \item \textbf{Daily modulation (first harmonic of sidereal frequency, $\omega_\oplus$):}
    \[
    C_{\omega_\oplus}(t) = 2 \cos \lambda \sin \lambda \cos \theta_X \sin \theta_X
    \cos(\omega_\oplus t - \phi_X).
    \]
    \item \textbf{Twice-daily modulation (second harmonic of sidereal frequency, $2\omega_\oplus$):}
    \[
    C_{2\omega_\oplus}(t) = \tfrac{1}{2}\cos^2\!\lambda\,\cos^2\!\theta_X
    \cos(2\omega_\oplus t - 2\phi_X).
    \]
\end{itemize}

The relative strengths of these components depend on the polarization latitude $\theta_X$ and the experimental latitude $\lambda$. 
When the polarization vector is parallel to the Earth's rotation axis ($\theta_X = \pm90^\circ$), the modulation vanishes because the projection $\cos^2\!\vartheta(t)$ remains constant. 
Similarly, for the zenith-aligned experiment conducted at either pole ($|\lambda| = 90^\circ$), the signal becomes time-independent for all polarization angles. 
The phase of the modulation is determined by the longitudinal difference between $\phi_X$ and the experiment’s location,as illustrated in Figure~\ref{fig:earth and modualtions}. Similar relations can be derived for the $\hat{\mathbf{V}}$.

\subsection{Expected Signal Lineshape}
\label{subsec:freqbin_response}

The converted dark-photon signal is not strictly monochromatic. Because the Galactic dark-matter halo has a three-dimensional rms speed $\sigma_v \simeq 270\,\mathrm{km/s}$ in the Galactic rest frame~\cite{HAYSTAC_procedure}, the corresponding photon frequency is broadened by $\Delta f / f_0 \sim \beta_2 \sim 10^{-6}$, where $\beta_2 = \sigma_v^2/c^2$. 
This intrinsic frequency spread defines the signal’s spectral lineshape, with rest-mass frequency $f_0 = m_{\rm DP} c^2/h$, and often described by the Standard Halo Model (SHM)~\cite{Lewin1996_SHM}. $S_{\rm SHM}(f; f_0)$ defines how the total converted power is distributed across a narrow spectral range around $f_0$.

The measured power spectrum is obtained from a discrete Fast Fourier Transform (FFT) of finite-duration time-domain data. Because no additional spectral window is applied, each frequency bin has the standard sinc-squared power response of a rectangular-window FFT. We denote the response of bin $j$, centered at $f_j$, to power at frequency $f$ by $B_j(f)$. Power at a given frequency can therefore be distributed across neighboring bins.

Convolving the intrinsic dark-matter lineshape with $B_j(f)$ gives the normalized bin weights   $\mathcal{C}_j(f_0)$, which describe how the expected signal power at trial frequency $f_0$ is distributed among neighboring bins.

\begin{equation}
\mathcal{C}_j(f_0)=
\frac{\displaystyle\int S_{\rm SHM}(f;f_0)\,B_j(f)\,df}
{\displaystyle\sum_{j'}\int S_{\rm SHM}(f;f_0)\,B_{j'}(f)\,df},
\label{eq:Cj_def}
\end{equation}

The normalization ensures that $\sum_j \mathcal{C}_j(f_0)=1$ over the signal window. For each trial frequency, the neighboring bins enter the likelihood jointly with a common signal amplitude $P_0$ and the corresponding weights $\mathcal{C}_j(f_0)$.

\section{Statistical Analysis}
\label{sec:analysis_likelihood}

The expected DP signal $\mathcal{S}_{kj}$, which incorporates the amplitude, time-dependent modulation, and frequency-bin response, serves as the foundation for all subsequent inferences.

This section describes how the likelihood function is constructed and applied under different statistical paradigms, including Bayesian and frequentist approaches. We then define the test statistics used to quantify the presence of a potential DP signal and finally derive the power upper limits. 

\subsection{Likelihood Framework}
\label{subsec:likelihood}

Although the true polarization angle is unknown, we can compute the expected modulation pattern $\mathcal{S}_{kj}(\chi,m_{\mathrm{DP}},\theta_X, \phi_X)$ for any given polarization direction and compare it with the experimentally measured time-dependent signal $y_{kj}$.

The probability of obtaining the data we observed, assuming a DP signal model with amplitude $P_0$, polarization orientation $(\theta_X, \phi_X)$, and mass $m_{\mathrm{DP}}$, can be described by the likelihood function as ~\cite{Cowan1998, PDG2024}
\begin{equation}
\mathcal{L}(D|P_0,m_{\mathrm{DP}},\theta_X,\phi_X) = \exp \left[-\frac{1}{2} \sum_{k,j}
\left( \frac{y_{kj} - \mathcal{S}_{kj}}{\sigma_{kj}} \right)^2\right],
\label{eq:likelihood_def}
\end{equation}
where $y_{kj}$ is the measured discrete signal power in the $j$-th frequency bin of the $k$-th spectrum, $\sigma_{kj}$ is the corresponding standard deviation from Gaussian noise.
This gives the log-likelihood  $\ln\mathcal{L} = -\tfrac{1}{2}\Lambda_{\rm data} + P_{0}\,\Lambda_{\rm cross} - \tfrac{1}{2}P_{0}^2\,\Lambda_{\rm temp}$, with:

\begin{align}
\Lambda_{\rm data} &= \sum_{k,j}\frac{y_{kj}^2}{\sigma_{kj}^2}, \\
\Lambda_{\rm cross} &= \sum_{j} \mathcal{C}_j(m_{\mathrm{DP}}) 
      \sum_{k} \frac{y_{kj}\,\eta_{kj}\,\cos^2\!\vartheta(\theta_X,\phi_X,t_k)}{\sigma_{kj}^2},\\
\Lambda_{\rm temp} &= \sum_{j} \mathcal{C}_j^2(m_{\mathrm{DP}}) 
      \sum_{k} \frac{\eta_{kj}^2\,\cos^4\!\vartheta(\theta_X,\phi_X,t_k)}{\sigma_{kj}^2}.
\end{align}
where $\Lambda_{\rm data}$ is the data-only contribution,
$\Lambda_{\rm cross}$ is the noise-weighted overlap between the data and the signal template,
and $\Lambda_{\rm temp}$ is the corresponding template self-overlap.

For any selected polarization angle and DP mass $(\theta_X,\phi_X,m_{\mathrm{DP}})$, the likelihood follows a Gaussian distribution in $P_{0}$, peaked at $\hat P_{0}$ with a width $\hat \sigma$ where:
\begin{equation}
\hat P_{0} = \frac{\Lambda_{\rm cross}}{\Lambda_{\rm temp}}, 
\qquad
\hat \sigma = \frac{1}{\sqrt{\Lambda_{\rm temp}}}.
\label{eq:P_hat}
\end{equation}
The local significance of an excess can then be estimated as
\begin{equation}
Z = \frac{\Lambda_{\rm cross}}{\sqrt{\Lambda_{\rm temp}}}.
\label{eq:significance}
\end{equation}
The quantities $\Lambda_{\rm cross}$ and $\Lambda_{\rm temp}$ are evaluated by summing over all spectra $k$ and frequency bins $j$ near $f_0$ (where $m_{\mathrm{DP}}$ peaks). We used $j\!\in\![j_0-1,\,j_0+4]$ to include six bins that capture the full dark-matter linewidth.

In practice, the evaluation of $\Lambda_{\rm cross}$ and $\Lambda_{\rm temp}$ proceeds on a discrete grid of parameters. 
The polarization angles $(\theta_X,\phi_X)$ are sampled uniformly on a linear grid, in our case, $21\times20$ points cover $[-\pi/2,\pi/2]\times[0,2\pi)$, and the signal power 
$P_0$ has a finer grid of 500 points from $-5 \times 10^{-20} \text{W}$ to $5 \times 10^{-20} \text{W}$, which covers about $\pm 6 \sigma$  of the averaged std. 
For each $(\theta_X,\phi_X)$ combination, the time-dependent projection $\cos^2\!\vartheta(\theta_X,\phi_X,t_k)$ correlates all spectra $k$.

$\Lambda_{\rm data}$ is independent of polarization and therefore needs to be evaluated only once over the full data set. In contrast, $\Lambda_{\rm cross}$ and $\Lambda_{\rm temp}$ contain the time-dependent factors $\cos^2\vartheta(t_k)$ and $\cos^4\vartheta(t_k)$, which couple the polarization hypothesis $(\theta_X,\phi_X)$ to the time series of spectra. A direct evaluation would require recomputing the sums over all timestamps $k$ for every polarization hypothesis and trial frequency, which becomes computationally expensive for $\sim10^4$ spectra and more than $10^6$ DP hypotheses. 

To accelerate the likelihood evaluation, we exploit the periodicity of the geometric factors. Since the time dependence arises purely from the Earth's rotation, the coupling terms can be exactly expanded into a finite series of sidereal harmonics. Specifically, $\cos^2\!\vartheta$ and $\cos^4\!\vartheta$ can be expanded in sidereal harmonics up to
 $n=2$ and $n=4$, respectively. This harmonic decomposition allows us to factorize the cross and template terms. For instance, $\Lambda_{\rm cross}(\theta_X,\phi_X,m_{\mathrm{DP}})$ can be rewritten as a linear combination of frequency-dependent, precomputable data sums and analytically derived angular coefficients:

\begin{equation}
\begin{split}
\Lambda_{\rm cross}
&= \sum_j \mathcal{C}_j(m_{\mathrm{DP}})
\sum_{n=0}^{2}
\Big[
b_{n,c}^{(2)}(\theta_X,\phi_X)\,
C_{n,j}^{\rm cross}
\\
&\quad +
b_{n,s}^{(2)}(\theta_X,\phi_X)\,
S_{n,j}^{\rm cross}
\Big]
\end{split}
\end{equation}
where the geometric coefficients $b_{n,c}^{(2)}$ and $b_{n,s}^{(2)}$ are analytical functions of the polarization angles $(\theta_X,\phi_X)$, independent of time and frequency. Here $n$ labels the sidereal harmonic, and $c$ and $s$ denote cosine and sine components. The superscripts $(2)$ and $(4)$ correspond to the expansions of $\cos^2\vartheta(t)$ and $\cos^4\vartheta(t)$, respectively. The polarization-independent harmonic data sums are defined as:

\begin{align}
C^{\rm cross}_{n,j} = &\sum_k \frac{y_{kj}\eta_{kj}}{\sigma_{kj}^2}\cos(n\omega_\oplus t_k),\\
S^{\rm cross}_{n,j} =  &\sum_k \frac{y_{kj}\eta_{kj}}{\sigma_{kj}^2}\sin(n\omega_\oplus t_k).
\end{align}

Similarly, $\Lambda_{\rm temp}(\theta_X,\phi_X,m_{\mathrm{DP}})$ could be written as:

\begin{equation}
    \Lambda_{\rm temp} = \sum_j \mathcal{C}_j^2 \sum_{n=0}^{4} \left[ b_{n,c}^{(4)}\,C_{n,j}^{\rm temp} + b_{n,s}^{(4)}\,S_{n,j}^{\rm temp} \right]
\end{equation}
with harmonic data sums defined as:
\begin{align}
C^{\mathrm{temp}}_{n,j} &= \sum_k \frac{\eta_{kj}^2}{\sigma_{kj}^2}\cos(n\omega_\oplus t_k),\\
S^{\mathrm{temp}}_{n,j} &= \sum_k \frac{\eta_{kj}^2}{\sigma_{kj}^2}\sin(n\omega_\oplus t_k).
\label{eq:harmonic_sums_temp}
\end{align}
This allows the sums to be factorized into precomputed harmonic components that are independent of the polarization angles, while the dependence on the polarization angles $(\theta_X,\phi_X)$ appears only through simple angular coefficients (detailed in Appendix \ref{subsec:Analytical Geometric Coefficients}).

After computing $\Lambda_{\rm cross}$ and $\Lambda_{\rm temp}$ for all sampled $(\theta_X,\phi_X)$, the likelihood $\mathcal{L}(P_{0},\theta_X,\phi_X,m_{\mathrm{DP}})$ of Eq.~(\ref{eq:likelihood_def}) is evaluated on the grid of $P_{0}$ values. 
 
Once $\mathcal{L}(D|P_{0},\theta_X,\phi_X,m_{\mathrm{DP}})$ is specified, the next step is to determine how to treat the unknown parameters $(\theta_X,\phi_X)$ to obtain the likelihood as a function of the physically interesting quantity $P_{0}$.

To deal with the nuisance parameters, we adopted both a Bayesian and a frequentist framework~\cite{PDG2024}.
At this stage, the likelihood depends on the full parameter vector $\psi \equiv (P_{0}, m_{\mathrm{DP}}, \nu)$, where $P_{0}$ is the parameter of interest and $\nu \equiv (\theta_X,\phi_X)$ denotes nuisance parameters describing the unknown polarization direction.
Both Bayesian and frequentist approaches start from the same likelihood $\mathcal{L}(D|\psi)$, but differ in how $\nu$ is treated.

\textbf{Bayesian approach:}
Bayesian inference combines the likelihood with priors via Bayes' theorem,
\begin{equation}
p(\psi|D)=\frac{\mathcal{L}(D|\psi)\,\pi(\psi)}
{\int \mathcal{L}(D|\psi)\,\pi(\psi)\,d\psi}\,
\label{eq:bayes_theorem}
\end{equation}
Here $D$ denotes the measured data, $\pi(\psi)$ is the prior density, and $p(\psi|D)$ is the posterior density. To obtain an inference that depends only on $P_{0}$ (for each  $m_{\mathrm{DP}}$), we marginalize over $\nu$ with a prior $\pi(\nu)$,
$\boldsymbol{\nu}=(\theta_X,\phi_X)$:
\begin{equation}
\mathcal{L}_{\rm Bayes}(D|P_{0},m_{\mathrm{DP}})
=\int \mathcal{L}(D|P_{0},\nu,m_{\mathrm{DP}})\,\pi(\nu)\,d\nu .
\label{eq:bayes_marg}
\end{equation}

\textbf{Frequentist approach:}
In our frequentist analysis, we treat the polarization angles as fixed but unknown nuisance parameters and profile over them by maximizing the likelihood ~\cite{Cowan2011,PDG2024}:
\begin{equation}
\mathcal{L}_{\mathrm{Freq}}(D|P_{0},m_{\mathrm{DP}})
   = \max_{\boldsymbol{\nu}}
   \mathcal{L}(D|P_{0},\boldsymbol{\nu},m_{\mathrm{DP}}).
\label{eq:freq_profile}
\end{equation}

\vspace{3pt}
\noindent
Equations~(\ref{eq:bayes_marg}) and (\ref{eq:freq_profile}) represent two general statistical paradigms that can be applied to any combination of data sets or experimental runs. 
These frameworks can be applied to the
GigaBREAD data, which consist of several independent sub-runs performed under different experimental conditions but sharing the same underlying signal model for a given DP mass $m_{\mathrm{DP}}$ and power
$P_{0}$.

In GigaBREAD, for each sub-run spectra were saved with a timestamp relative to the start time of the data taking run, so the relative phase of the sidereal modulation could be determined within that run. 
However, absolute timestamps in \textit{GigaBREAD2023} were not recorded, meaning the global modulation phase could not be aligned across sub-runs. Because the setup is fixed in the earth reference frame, this lack of absolute timing only obscures the $\phi_X$ information across runs, leaving $\theta_X$ intact.
Consequently, the azimuthal polarization angle $\phi_X$  was treated as an independent nuisance parameter for each run: it was marginalized (separately summed) within each run in the Bayesian combination and profiled within each run in the frequentist case.
The overall product across runs is therefore taken after the inner summation or maximization over~$\phi_X$.
By contrast, the polarization latitude~$\theta_X$ represents a global direction of the DP field that is common to all runs, and is therefore summed or profiled outside the run-level product. 

\begin{equation}
\begin{split}
\mathcal{L}_{\mathrm{Bayes}}(D&|m_{\mathrm{DP}},P_{0}) \\
   &=\sum_{\theta_X}w_{\theta_X}\prod_n\left[\frac{1}{N_\phi}\sum_{\phi_X}\mathcal{L}_n(D|m_{\mathrm{DP}},\theta_X,\phi_X,P_0)\right],
\end{split}
\end{equation}
\begin{equation}
\mathcal{L}_{\mathrm{Freq}}(D|m_{\mathrm{DP}},P_{0})
   = \max_{\theta_X}\,
   \prod_{n}\max_{\phi_X}
   \mathcal{L}_{n}(D|m_{\mathrm{DP}},\theta_X,\phi_X,P_{0}),
\label{eq:bayes_freq_practical}
\end{equation}
where the indices $n$ label different data runs, and the summations or maximizations represent the Bayesian marginalization and frequentist profiling over the nuisance parameters $(\theta_X,\phi_X)$.
The ordering of the operations-a sum over $\theta_X$ outside the product over runs, followed by an internal sum (or maximum) over $\phi_X$ within each run-reflects the practical data structure of the experiment. We adopt a uniform prior over solid angle $w_{\theta_X} \propto \cos\theta_X $, so that any direction has equal probability.

\subsection{Bayes Factors and Test Statistics for Signal Detection}
\label{subsec:teststat}

To evaluate whether the data contain a statistically significant excess consistent with a signal, we construct a Log-Bayesian factor $k_0$  and a frequentist test statistic $q_0$ that quantifies the incompatibility between the data and the null hypothesis. 
\begin{figure*}[htbp]
\centering
\includegraphics[width=0.49\textwidth]{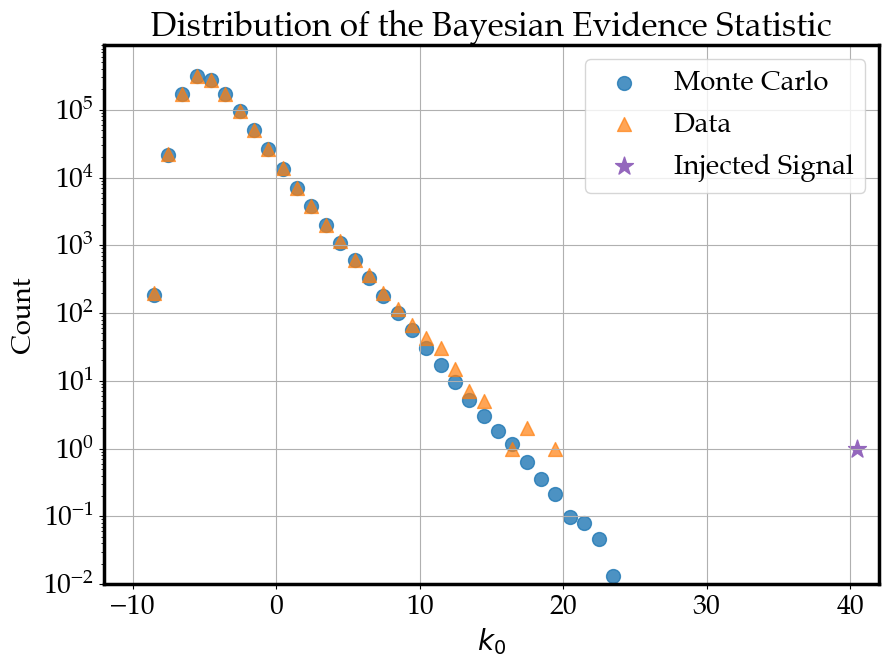}
\includegraphics[width=0.49\textwidth]{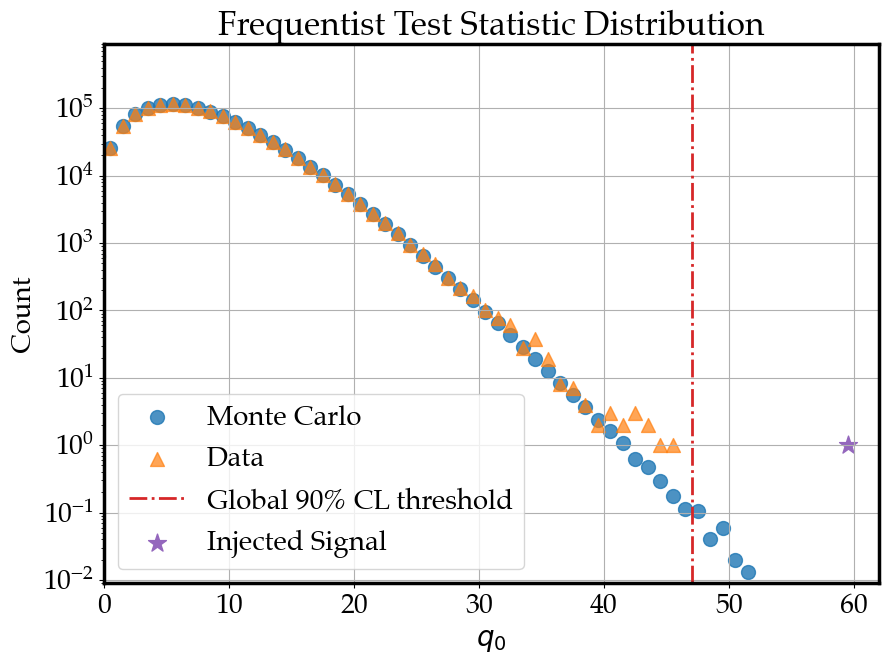}
\caption{Distributions of the Bayesian evidence statistic $k_0$ (left) and frequentist test statistic $q_0$ (right) for the experimental data and background-only Monte Carlo simulations. The Monte Carlo histogram gives the background-only null distribution.   The injected-signal point shown here is a leakage-induced excess in a relatively distant neighboring bin, with both $k_0$ and $q_0$ exceeding 100 near the signal center, beyond the plotted ranges. frequentist global 90\% CL threshold is calibrated from the Monte Carlo trials;}
\label{fig:teststat_distributions}
\end{figure*}

\textbf{Bayes Factor:}
In the Bayesian formulation~\cite{Sivia2006,PDG2024}, nuisance parameters are marginalized over. To quantify the support for a DP signal at each mass, we evaluate the local Bayes factor $\mathcal{B}_{10}$, which compares the marginalized likelihood of the signal-plus-background hypothesis $H_1$ ($0\leq P_0\leq P_{\max}$) with that of the background-only hypothesis $H_0$ ($P_0=0$). We report the corresponding Bayesian evidence statistic $k_0\equiv2\ln\mathcal{B}_{10}$:
\begin{equation}
k_0 = 2\ln \frac{L(D|H_1)}{L(D|H_0)} = 2\ln \frac{\displaystyle\int_{0}^{P_{\max}} \mathcal{L}(P_{0})\pi(P_{0})\,dP_{0}}{\mathcal{L}(0)},
\label{eq:q0_bayes}
\end{equation}
where $\pi(P_{0})=1/P_{\max}$ for $0\leq P_0\leq P_{\max}$ and zero otherwise. Under the background-only hypothesis, the evidence is simply $\mathcal{L}(P_0=0)$.

\textbf{Frequentist test statistic:}

In the frequentist approach, the detection statistic compares the best-fit likelihoods in signal-plus-background ($P_{0} > 0$) and the restricted null hypothesis ($P_{0} = 0$)~\cite{Cowan2011,PDG2024}. We restrict to the positive power so that downward background fluctuations yield zero discovery significance:
\begin{equation}
q_0 = 2\ln\frac{\displaystyle\sup_{P_{0}\ge0}\mathcal{L}(P_{0})}{\mathcal{L}(0)}
\label{eq:q0_freq}
\end{equation}

Here $\displaystyle\sup_{P_{0}\ge0}\mathcal{L}(P_{0})$ is the maximum likelihood in a non-negative power range.

\textbf{Application to GigaBREAD:}
Both Bayesian and frequentist analyses are applied to the GigaBREAD data using the corresponding likelihoods $\mathcal{L}_{\mathrm{Bayes}}(P_{0},m_{\mathrm{DP}})$ and $\mathcal{L}_{\mathrm{Freq}}(P_{0},m_{\mathrm{DP}})$, from which the Log-Bayes factor $k_0(m_{\mathrm{DP}})$ and test statistics $q_0(m_{\mathrm{DP}})$ are computed for each DP mass. 

For a Gaussian, nuisance-free likelihood, the test statistic would follow the Chernoff distribution ~\cite{Cowan2011}. However, our likelihood includes time-dependent modulation and polarization profiling, with $\phi$ treated as an independent nuisance parameter per run, leading to a \emph{non-Wilks} distribution that must be calibrated empirically. To validate these results and accurately characterize the background, we generated toy Monte Carlo ensembles by adding normally distributed noise to the real data baseline(see Appendix~\ref{sec:appendix_mc}).

Three hundred independent noise toy ensembles—each equivalent in size to the full dataset—were produced, corresponding to a total of $1{,}151{,}965\times 300$ test-statistic values. Each toy realization underwent the same data-processing pipeline as the experiment. The resulting $k_0$ and $q_0$ histograms were averaged to form a smooth empirical null distribution $f(k_0|H_0)$ and  $f(q_0|H_0)$ with improved statistical stability and extended tail coverage. The averaged curve, shown in Fig.~\ref{fig:teststat_distributions}, closely matches the data histogram while exhibiting a better resolved tail. Although Bayesian discovery evidence does not rely on null MC, the $f(k_0|H_0)$ distribution gives a good insight of how the Bayes factor behaves under purely background-driven fluctuations. Meanwhile, the $f(q_0|H_0)$ distribution is used to define the frequentist global significance thresholds discussed below.

\textbf{Global Bayesian Evidence}:

In the Bayesian framework, since the true DP mass is unknown, we must marginalize the evidence over the entire search range. Assuming a uniform prior across the $N = 1{,}151{,}965$ scanned mass points, the global Bayesian evidence statistic $k_0^{\mathrm{global}}$ is obtained by averaging the local Bayes factors over the scanned mass points:

\begin{figure}[tb]
  \centering
  \includegraphics[width=0.45\textwidth]{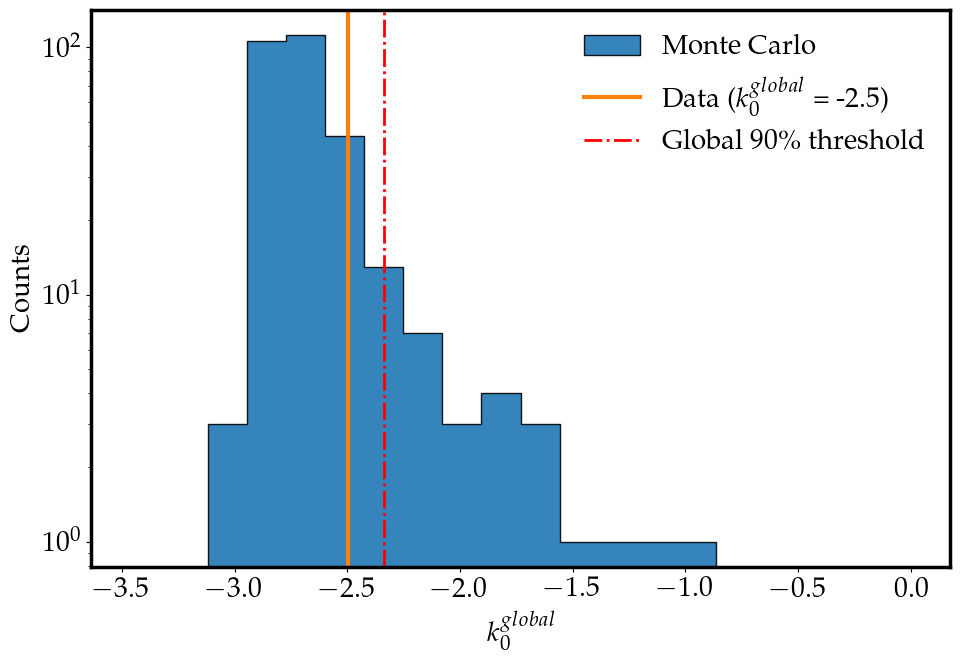}
  \caption{The observed Bayesian evidence statistic of the experimental data ($k_0^{\mathrm{global}} = -2.5$) falls within the background-only Monte Carlo distribution and below the candidate threshold of -2.3, indicating no signal detection. Test data with unmasked injected signals yields a $k_0^{\mathrm{global}}$ of approximately 189 (far beyond the plotted axis limits).}
  \label{fig:Global Bayes}
\end{figure}

\begin{equation}
k_0^{\mathrm{global}} = 2\ln \left( \frac{1}{N} \sum_{i=1}^{N} \frac{L(D|H_1, m_{\mathrm{DP}, i})}{L(D|H_0,m_{\mathrm{DP},i})} \right),
\label{eq:k0_global}
\end{equation}

where $L(D|H_1, m_{\mathrm{DP}, i})$ is the marginalized likelihood for a specific mass evaluated in Eq.~(\ref{eq:q0_bayes}), and $L(D|H_0) = \mathcal{L}(P_0 = 0)$ is the background likelihood.

We use the background-only MC distribution to define a global candidate threshold for this statistic. With 300 background-only MC toy datasets, we obtain a global 90\% threshold of $k_{0,\mathrm{th}}^{\mathrm{global}}=-2.3$, as shown in Fig.~\ref{fig:Global Bayes}. Values above this threshold are flagged as signal candidates, while lower values are consistent with the background-only hypothesis.

\textbf{Global Frequentist Threshold:}
In the frequentist framework, rather than assuming a fixed number of independent trials, we estimate an effective trials factor $N_{\mathrm{eff}}$ directly from Monte Carlo (null) realizations.

Specifically, for each null realization we compute the maximum test statistic over the full mass range, and compare the cumulative distributions of the local statistic $q$ and the scan maximum $q_{\max}$. The effective number of trials is defined through
\[
F_{\max}(q) \approx F_{\mathrm{local}}(q)^{N_{\mathrm{eff}}},
\]
where $F_{\mathrm{local}}$ and $F_{\max}$ are the cumulative distribution functions of the local statistic and the scan maximum, respectively. 
From this relation we extract
\[
N_{\mathrm{eff}} = \frac{\ln F_{\max}(q)}{\ln F_{\mathrm{local}}(q)},
\]
evaluated at the median of the null scan-maximum distribution.
\[
N_{\mathrm{eff}} \simeq 5 \times 10^5,
\]
which lies between the number of frequency bins ($N_j = 230,392$)and the total number of scanned templates ($N_{m_{DP}} = 1{,}151{,}965$), reflecting correlations between nearby frequencies and sub-bin offsets.

Given $N_{\mathrm{eff}}$, the relation between local and global probabilities is
\[
p_{\mathrm{global}} = 1 - (1 - p_{\mathrm{local}})^{N_{\mathrm{eff}}},
\]

The global $90\%$ candidate threshold is obtained from the local null-MC distribution by using $N_{\mathrm{eff}}$ to convert a $10\%$ global background-only exceedance probability to the corresponding local probability (Fig.~\ref{fig:teststat_distributions}). A scan whose maximum $q_0$ exceeds this threshold is flagged as a signal candidate.

\subsection{Power-Limit Estimation}
\label{subsec:powerlimit}

No tested DP mass exceeds the global detection threshold in Sec.~\ref{subsec:teststat}, the data are interpreted as consistent with the background-only hypothesis. 
We therefore set upper limits on the DP conversion power $P_{0}$ as a function of DP mass~\cite{PDG2024}.

\textbf{Bayesian Approach:} 
Assuming a uniform prior for physically allowed signal powers ($P_0 \ge 0$), the 90\% credible upper limit $P_{90\%}^{\mathrm{Baye}}$ is derived by integrating the marginalized posterior probability distribution:
\begin{equation}
    \frac{\int_{0}^{P_{90\%}^{\mathrm{Baye}}} \mathcal{L}(P_0) \, dP_0}{\int_{0}^{\infty} \mathcal{L}(P_0) \, dP_0} = 0.90
\end{equation}

\textbf{Frequentist Approach:} 
We use a Monte Carlo-calibrated Neyman construction to obtain a local frequentist upper limit, which is designed to yield an upper limit at or above the true signal power $P_0$ in at least $90\%$ of repeated experiments. Specifically, we evaluate the test statistic $q0$ (Eq~\ref{eq:q0_freq}) across a range of hypothesized dark photon powers($P_0 = P_{test}$) and polarization directions.   
To determine the critical threshold $q_{\mathrm{cut}}(P_{\mathrm{test}})$ for a specific signal hypothesis, we perform the following construction:
\begin{enumerate}
    \item \textbf{Simulate:} For each test power $P_{\mathrm{test}}$, we generate extensive Toy Monte Carlo datasets $D_{\mathrm{toy}}$ spanning various DP polarization directions.
    \item \textbf{Worst-Case Distribution:} We evaluate the test statistic $q_0(D_{\mathrm{toy}})$ and construct its empirical probability distribution for each tested polarization direction. We then select the distribution with the smallest lower 10\% quantile as the worst-case distribution.
    \item \textbf{Threshold:} We define the critical threshold $q_{\mathrm{cut}}(P_{\mathrm{test}})$ as the lower 10\% quantile of this worst-case $q_{\mathrm{0}}$ distribution.
\end{enumerate}
To prevent unphysical negative limits arising from downward background fluctuations, we floor the inverted upper limit at the background-only hypothesis, $P_0=0$. Finally, the 90\% confidence-level upper limit $P_{90\%}^{\mathrm{Freq}}$ is determined by the Neyman inversion boundary $q_{\mathrm{obs}}=q_{\mathrm{cut}}(P_{\mathrm{test}})$, with test powers satisfying $q_{\mathrm{obs}}<q_{\mathrm{cut}}(P_{\mathrm{test}})$ excluded. Coverage tests confirm that calibration against the worst-case scenario provides at least 90\% empirical coverage over the polarization directions and injected signal powers tested.

\section{Search Results}
\label{sec:search_results}

We find no statistically significant evidence for a dark photon signal within the surveyed mass range of 44--52~$\mu\mathrm{eV}$. Consequently, we proceed to set upper limits on the signal parameters.

\subsection{Dark-Photon Power Limits}
\label{subsec:DP_limits}

\begin{figure}[tb]
  \centering
  \includegraphics[width=0.5\textwidth]{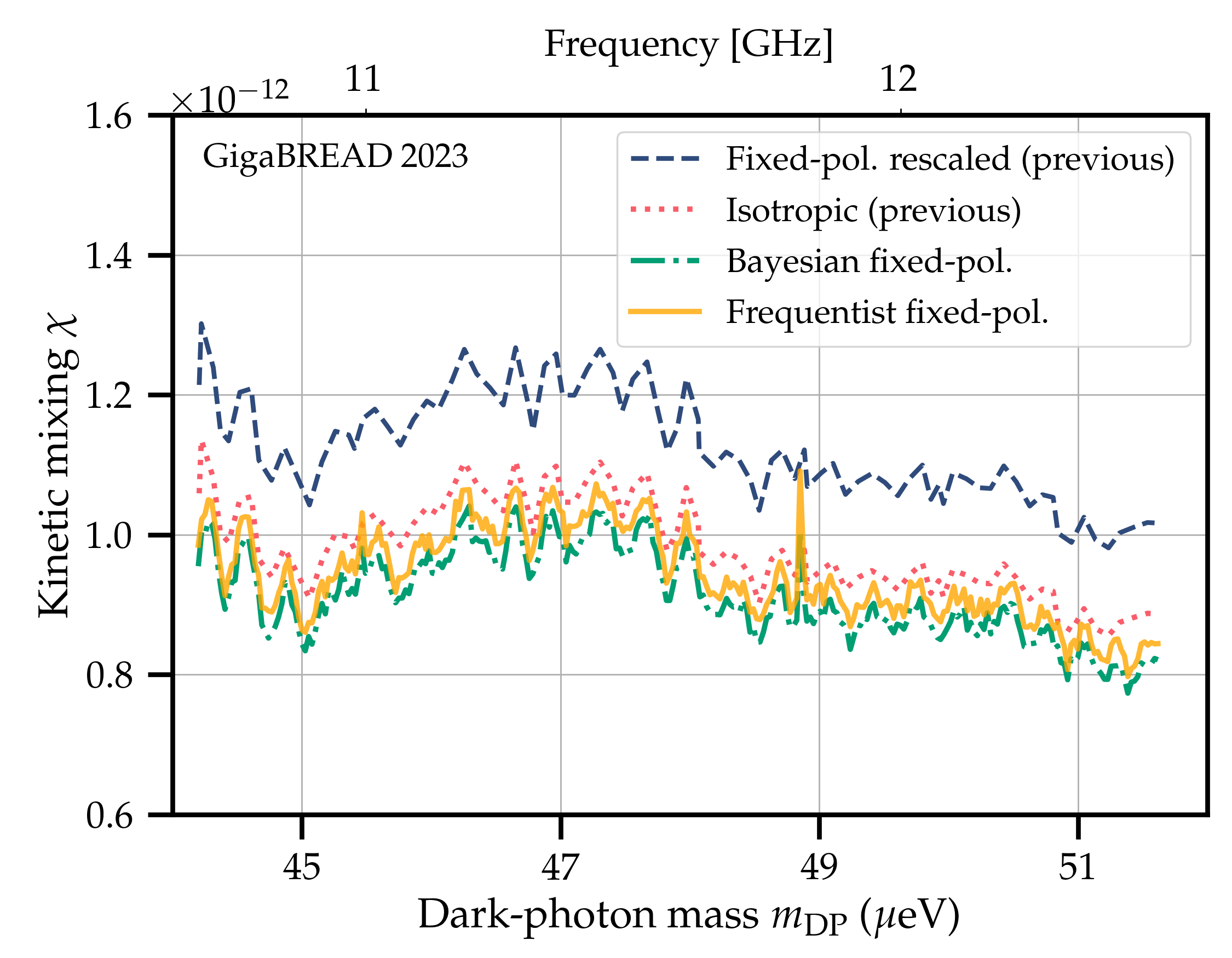}
  \caption{
  Comparison of 90\% exclusion limits on the kinetic-mixing parameter~$\chi$ obtained from different analyses of the GigaBREAD2023 data. The isotropic curve shows the published GigaBREAD DP limit assuming $\langle\cos^2\theta\rangle=1/3$ for random polarization factor~\cite{GigaBREAD2024_PRL}. The conservative curve shows the previous fixed-polarization rescaling used in the \textsc{AxionLimits} repository~\cite{AxionLimits}. The two lower curves show the Bayesian and frequentist results from the polarization-aware analysis of the same GigaBREAD 2023 dataset. For the plotted curves, we combine distributions from neighboring trial frequencies before extracting the upper limits.}
  \label{fig:DP_limit_comparison}
\end{figure}

Following the likelihood analysis in Section~\ref{subsec:powerlimit}, we first establish the $90 \%$ upper limit on the signal power $P_0$. We increase the power upper limits by the frequency-dependent systematic uncertainty (~$23 \%$) from Ref.~\cite{GigaBREAD2024_PRL}. From this adjusted power limit, we subsequently derive the $90\%$ exclusion limits on the kinetic-mixing parameter $\chi$ as a function of the DP mass $m_{\mathrm{DP}}$. Figure~\ref{fig:DP_limit_comparison} compares the updated GigaBREAD limits with previously published results.

\begin{figure*}[htbp]
    \centering
    \includegraphics[width=0.99\textwidth]{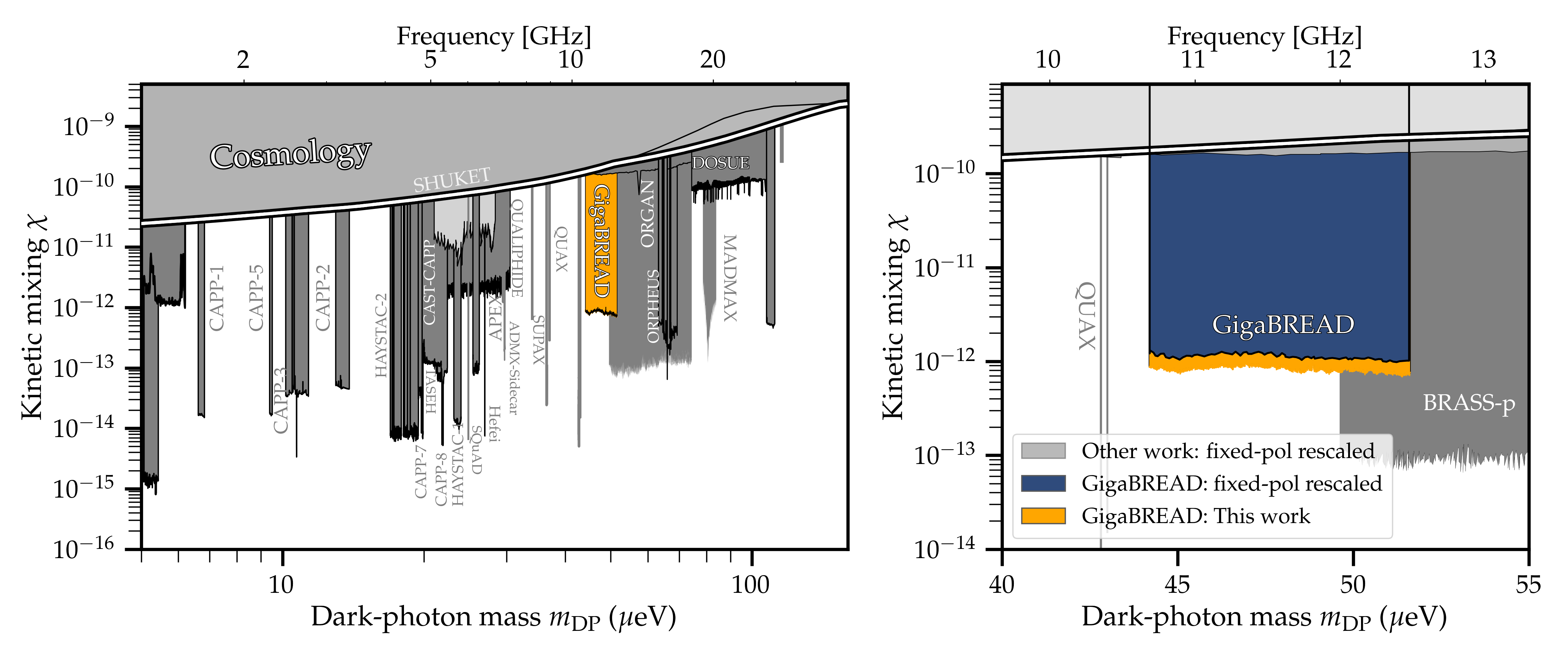}
    \caption{
    \textbf{Left:} Global fixed polarization DP parameter-space limits in the kinetic mixing~$\chi$ versus mass~$m_{\mathrm{DP}}$ plane. Gray shaded regions indicate existing exclusions from cosmology, haloscope, and dish-antenna experiments (SHUKET, QUALIPHIDE, BRASS-p, DOSUE, etc.), taken from AxionLimits \cite{AxionLimits}. The orange region shows the updated limit derived in this work. \textbf{Right:} Magnified view of the 40–55\,$\mu$eV region containing the GigaBREAD search range. The blue region denotes the previous rescaled fixed-polarization reinterpretation of the GigaBREAD exclusion limit. The orange region shows the new fixed-polarization exclusion limit obtained in this work.}
    \label{fig:DP_limits}
\end{figure*}

The isotropic result represents the original GigaBREAD2023 limit reported in Ref.~\cite{GigaBREAD2024_PRL}, which assumed random polarization and therefore used the averaging factor $\langle\cos^2\theta\rangle=1/3$ The top blue curve in Figure~\ref{fig:DP_limit_comparison} shows a previous fixed-polarization rescale of the original limit, obtained by applying the polarization factor $\langle\cos^2\theta\rangle_T=0.254$ on time averaged data over the observation duration $T$, following the implementation in AxionLimits~\cite{AxionLimits}. The two lower curves show the fixed-polarization limits obtained by applying our Bayesian and frequentist methods to the same GigaBREAD 2023 dataset. These limits are about 20\% lower in $\chi$ than the previously rescaled limit and are comparable to the original isotropic limit. The improvement can be understood in terms of the additional modulation information contained in a modulating signal versus a time-independent signal.

\subsection{Comparison with Previous Experiments}
\label{subsec:comparison}

Figure~\ref{fig:DP_limits} shows the updated fixed-polarization limits from the combined GigaBREAD 2023 and 2024 datasets, together with previous constraints.

The left panel summarizes constraints on the kinetic-mixing parameter $\chi$ as a function of the DP mass $m_{\mathrm{DP}}$. The gray shaded regions correspond to previous laboratory and astrophysical exclusions, including cosmological bounds at high~$\chi$ and limits from resonant-cavity haloscopes such as ADMX~\cite{ADMX2021}, HAYSTAC~\cite{HAYSTAC2023}, ORGAN~\cite{ORGAN_2023,ORGAN2017}, and QUAX~\cite{QUAX2023}. The broadband dish-antenna constraints from SHUKET~\cite{SHUKET2019}, QUALIPHIDE~\cite{QUALIPHIDE2023}, BRASS-p~\cite{BRASS-p2023}, and DOSUE~\cite{DOSUE2023} are also shown for comparison.

The right panel in Figure~\ref{fig:DP_limits} shows a magnified version of the 44–52\,$\mu$eV (10.7--12.5\,GHz) range covered by the GigaBREAD measurements. For comparison, the dark blue region denotes the previous rescaled fixed-polarization reinterpretation of the GigaBREAD exclusion limit, and the orange region shows fixed-polarization exclusion limit derived in this work, taking the larger of the Bayesian and frequentist upper limits at each mass. This result at $\chi \sim 10^{-12}$ is slightly stronger than the previous limit assuming random DP polarization~\cite{GigaBREAD2024_PRL}, benefiting from additional data collected with a different detector orientation.

\subsection{Implications for Detector Design}
\label{subsec:detector_design}

\begin{table}[htbp]
\centering
\begin{tabular}{lcc}
\hline\hline
Configuration & Time-averaged & Modulation-matched \\
\hline
Zenith axis &
$|\lambda|=35.3^\circ$ &
$|\lambda|=38.0^\circ$ \\
&
$\langle\cos^2\theta\rangle_T=0.333$ &
$\sqrt{\langle\cos^4\theta\rangle_T}=0.380$ \\
\hline
Horizontal axis &
$|\cos\lambda\cos\phi_d|=1/\sqrt{3}$ &
$|\cos\lambda\cos\phi_d|=0.616$ \\
&
$\langle\cos^2\theta\rangle_T=0.333$ &
$\sqrt{\langle\cos^4\theta\rangle_T}=0.380$ \\
\hline\hline
\end{tabular}
\caption{Optimal detector geometries for observations spanning an integer number of sidereal days, assuming uniform time weighting, obtained by maximizing the minimum response over fixed DP polarization directions. Here $\lambda$ is the geographic latitude and $\phi_d$ is the horizontal detector-axis orientation measured from north.}
\label{tab:optimal_geometry}
\end{table}
For a fixed but unknown DP polarization, we require the frequentist limit to maintain coverage for every tested polarization direction; therefore, we evaluate each detector configuration using its least-sensitive direction. In the time-averaged case, the accumulated response is characterized by $\langle\cos^2\vartheta(t)\rangle_T$. As the Earth rotates, the detector axis sweeps out a cone, and the optimal cone gives a maximally symmetric response: every fixed polarization direction has the same time-averaged response, thereby maximizing sensitivity to the least-sensitive polarization. For a zenith-pointing detector, this occurs at a half-opening angle of $54.7^\circ$ ($|\lambda|=35.3^\circ$) and gives $\langle\cos^2\vartheta(t)\rangle_T=1/3$, consistent with Ref.~\cite{Caputo2021}. In the modulation-matched likelihood used here, however, $\Lambda_{\rm temp}\propto\sum_i\cos^4\vartheta(t_i)$, so the corresponding polarization factor is $\sqrt{\langle\cos^4\vartheta(t)\rangle_T}$. This gives more weight to times when the detector is closely aligned with the DP polarization, improving sensitivity compared with time averaging alone. As a result, the optimal half-opening angle decreases from $54.7^\circ$ to approximately $52.0^\circ$ ($|\lambda|=38.0^\circ$). The corresponding zenith and horizontal-axis geometries are summarized in Table~\ref{tab:optimal_geometry}.

\section{Conclusion}
\label{subsec:summary_outlook}

Vector dark matter produced by a misalignment mechanism may retain an approximately fixed polarization, leading to a time-dependent signal modulation due to Earth's rotation. However, previous GigaBREAD analysis and several major dish-antenna searches have assumed randomly polarized dark photon dark matter.

In this work, we reanalyze the GigaBREAD data using a polarization-matched likelihood framework that incorporates the fixed-polarization geometry and its time-dependent projection onto the detector. We find no evidence for a dark photon signal and set fixed-polarization limits on the kinetic mixing parameter $\chi \sim 10^{-12}$ over 44–52\,$\mu$eV mass range. These limits are about $20\%$ lower in $ \chi $ than a naive limit obtained by rescaling the previous GigaBREAD limit under the fixed-polarization assumption.

Broadband dish antennas can observe the full bandwidth of the experiment for the duration of the experimental run, making them particularly suitable for this type of analysis. Our results demonstrate that incorporating polarization modulation and likelihood-based approaches can improve the sensitivity of broadband experiments to polarized dark photons.

Future extensions of this analysis will explore additional data sets and further refinements of the statistical framework. The methodology developed here can be directly applied to other broadband dark-photon searches and similar experiments involving time-dependent or polarization-dependent signals.

\section*{Acknowledgments}
We acknowledge Abigail Vieregg and her group for providing access to the RF-shielded room and RF absorber materials. We thank Argonne National Laboratory, operated under Contract No. DE-AC02-06CH11357, for use of the solenoid magnet facility, which enabled the search for axion-like particles. We are grateful to Dan Zhang and the ADMX Collaboration for helpful discussions about the analysis, and to all BREAD collaborators for inspiring discussions. This work was funded by the U.S. Department of Energy through the resources of the Fermi National Accelerator Laboratory (Fermilab), a DOE Office of Science HEP User Facility. Fermilab is managed by Fermi Forward Discovery Group, LLC (FermiForward) under Contract No. 89243024CSC000002. J.Y. was supported by the Fermilab LDRD program under FNAL-LDRD-2021-047 and by the Illinois Institute of Technology. S.K. is supported by the Faculty of Arts and Sciences of Harvard University.

Generative AI Disclosure: ChatGPT-5.6 was used to check grammar and for style editing. All scientific content and calculations were created and reviewed for accuracy by the authors, who also verified the references and take full responsibility for the manuscript.

\appendix
\section{Monte Carlo Simulation and Null Hypothesis Characterization}
\label{sec:appendix_mc}

To generate background-only toy datasets ($H_0$), we employ an Artifact-Isolation Monte Carlo method. This approach models the correlations between neighboring frequency bins and the reduction in residual variance introduced by Savitzky--Golay baseline subtraction, without reproducing residual structure caused by imperfect fits to the instrumental baseline.

The simulated background data products are constructed by isolating these statistical artifacts and scaling them by the empirically derived baseline. For an ideal standard normal noise field $\mathcal{G}_{kj} \sim \mathcal{N}(0,1)$, the filtered noise residual $z^{\rm H_0}_{kj}$ is defined by subtracting the filter's absorption response:
\begin{equation}
    z^{\rm H_0}_{kj} = \mathcal{G}_{kj} - \text{SavgolFilter}(\mathcal{G}_{kj}).
\end{equation}
The final simulated $H_0$ dataset is then assembled by the measured noise standard deviation:
\begin{align}
    y^{\rm H_0}_{kj} &= z^{\rm H_0}_{kj} \cdot \sigma^{\text{real}}_{kj}, \\
    \sigma^{\rm H_0}_{kj} &= \sigma^{\text{real}}_{kj}.
\end{align}
This construction ensures the whitened Toy residuals, $y^{\rm H_0}_{kj} / \sigma^{\rm H_0}_{kj}$, model the filter-induced correlations and variance suppression while remaining free of deterministic structural biases.
\section{Simplifying the Likelihood Calculation}
\label{sec:appendix_math}

Evaluating the polarization-resolved likelihood across a multi-dimensional parameter grid $(\theta_X, \phi_X, P_0, f_0)$ introduces a prohibitive computational bottleneck $\mathcal{O}(N_k)$ if the sum over the $\sim 10^4$ temporal spectra is performed explicitly. We decouple the time-dependent modulations from the polarization parameters through an analytical harmonic factorization optimized for rapid evaluation.

\subsection{Harmonic Factorization of the Likelihood}
First, to remove explicit dependence on the slowly varying detection efficiency $\eta_{kj}$, we normalize the measured signal and its standard deviation: $y'_{kj}=y_{kj}/\eta_{kj}$ and $\sigma'_{kj}=\sigma_{kj}/\eta_{kj}$. 

The modulation operators $\cos^2\!\vartheta_k=|\hat{X}\!\cdot\!\hat{\mathbf D}(t_k)|^2$ and $\cos^4\!\vartheta_k$ depend on the sidereal rotation $\omega_\oplus t_k$ and the dark-photon polarization angles $(\theta_X,\phi_X)$. These projections can be exactly expanded as finite Fourier series up to order $m=2$ and $m=4$, respectively:
\begin{equation}
\begin{split}
|\hat{X}\!\cdot\!\hat{\mathbf D}(t_k)|^m = \sum_{n=0}^{m} \Big[\,&b_{n,c}^{(m)}(\theta_X,\phi_X)\cos(n\omega_\oplus t_k) \\
&+ b_{n,s}^{(m)}(\theta_X,\phi_X)\sin(n\omega_\oplus t_k)\,\Big],
\end{split}
\label{eq:harmonic_expansion}
\end{equation}
where the coefficients $b_{n,c}^{(m)}$ and $b_{n,s}^{(m)}$ are analytical functions of the detector geometry and polarization angles.

Substituting Eq.~\eqref{eq:harmonic_expansion} into the likelihood allows us to separate the time summations from the parameter scan. For each frequency bin $j$, we precompute the polarization-independent time-domain harmonic sums:
\begin{align}
C^{\rm cross}_{n,j} &= \sum_k \frac{y'_{kj}}{{\sigma'_{kj}}^2}\cos(n\omega_\oplus t_k), \\
S^{\rm cross}_{n,j} &= \sum_k \frac{y'_{kj}}{{\sigma'_{kj}}^2}\sin(n\omega_\oplus t_k), \label{eq:C2S2}\\[3pt]
C^{\rm temp}_{n,j} &= \sum_k \frac{1}{{\sigma'_{kj}}^2}\cos(n\omega_\oplus t_k), \\
S^{\rm temp}_{n,j} &= \sum_k \frac{1}{{\sigma'_{kj}}^2}\sin(n\omega_\oplus t_k). \label{eq:C4S4}
\end{align}

The polarization-dependent terms $\Lambda_{\rm cross}$ and $\Lambda_{\rm temp}$ then collapse into direct vector contractions between the static geometric coefficients and the precomputed temporal harmonics, summed over the relevant frequency bins with their spectral weights $\mathcal{C}_j(f_0)$:
\begin{align}
\Lambda_{\rm cross} &= \sum_{j} \mathcal{C}_j(f_0) \sum_{n=0}^{2} \Big[\,b_{n,c}^{(2)}\,C^{\rm cross}_{n,j} + b_{n,s}^{(2)}\,S^{\rm cross}_{n,j}\,\Big], \\
\Lambda_{\rm temp} &= \sum_{j} \mathcal{C}_j^2(f_0) \sum_{n=0}^{4} \Big[\,b_{n,c}^{(4)}\,C^{\rm temp}_{n,j} + b_{n,s}^{(4)}\,S^{\rm temp}_{n,j}\,\Big].
\end{align}
After precomputing the harmonic sums, each likelihood evaluation uses a fixed number of terms rather than summing over all $N_k$ spectra, without changing the likelihood.
\subsection{Analytical Geometric Coefficients}
\label{subsec:Analytical Geometric Coefficients}
The angular amplitudes $b_{n,c}^{(m)}(\theta_X,\phi_X)$ and $b_{n,s}^{(m)}(\theta_X,\phi_X)$ are uniquely determined by the detector's geographical orientation. The non-zero coefficients for the geometries analyzed in this work are given below.

\subsubsection*{1. Zenith-aligned configuration }
At latitude $\lambda$, defining $A_0=\sin\theta_X\sin\lambda$ and $A_1=\cos\theta_X\cos\lambda$:
\begin{align}
b_{0,c}^{(2)} &= A_0^2+\tfrac12A_1^2, \\
b_{1,c}^{(2)} &= 2A_0A_1\cos\phi_X, \\
b_{1,s}^{(2)} &= 2A_0A_1\sin\phi_X, \\
b_{2,c}^{(2)} &= \tfrac12A_1^2\cos2\phi_X, \\
b_{2,s}^{(2)} &= \tfrac12A_1^2\sin2\phi_X.
\end{align}
The $m=4$ coefficients are obtained similarly by expanding $(|\hat{X}\!\cdot\!\hat{Z}|^2)^2$:
\begin{align}
b_{0,c}^{(4)} &= A_0^4+\tfrac{3}{8}A_1^4+3A_0^2A_1^2, \\
b_{1,c}^{(4)} &= (4A_0^3A_1+3A_0A_1^3)\cos\phi_X, \\
b_{1,s}^{(4)} &= (4A_0^3A_1+3A_0A_1^3)\sin\phi_X, \\
b_{2,c}^{(4)} &= (3A_0^2A_1^2+\tfrac12A_1^4)\cos2\phi_X, \\
b_{2,s}^{(4)} &= (3A_0^2A_1^2+\tfrac12A_1^4)\sin2\phi_X, \\
b_{3,c}^{(4)} &= A_0A_1^3\cos3\phi_X, \\
b_{3,s}^{(4)} &= A_0A_1^3\sin3\phi_X, \\
b_{4,c}^{(4)} &= \tfrac18A_1^4\cos4\phi_X, \\
b_{4,s}^{(4)} &= \tfrac18A_1^4\sin4\phi_X.
\end{align}

\subsubsection*{2. Horizontal-axis configuration}
For a horizontal detector with azimuth $\phi_d$ relative to North, defining $A_0 = -\sin\theta_X\cos\lambda\cos\phi_d$, $A_{1c} = \cos\phi_d\cos\theta_X\sin\lambda$, and $A_{1s} = \sin\phi_d\cos\theta_X$:
\begin{align}
b_{0,c}^{(2)} &= A_0^2+\tfrac12(A_{1c}^2+A_{1s}^2), \\
b_{1,c}^{(2)} &= 2A_0A_{1c}\cos\phi_X - 2A_0A_{1s}\sin\phi_X, \\
b_{1,s}^{(2)} &= 2A_0A_{1c}\sin\phi_X + 2A_0A_{1s}\cos\phi_X, \\
b_{2,c}^{(2)} &= \tfrac12(A_{1c}^2-A_{1s}^2)\cos2\phi_X \notag\\
&\quad - A_{1c}A_{1s}\sin2\phi_X, \\
b_{2,s}^{(2)} &= \tfrac12(A_{1c}^2-A_{1s}^2)\sin2\phi_X \notag\\
&\quad + A_{1c}A_{1s}\cos2\phi_X.
\end{align}
The fourth-order coefficients for the horizontal configuration are obtained by squaring the second-order expansion and collecting terms at each sidereal harmonic.
\bibliographystyle{unsrt}
\bibliography{references}

@article{ArgonneAxion2025,
  title = {First Axionlike Particle Results from a Broadband Search for Wavelike Dark Matter in the 44 to $52\text{ }\text{ }\mathrm{\ensuremath{\mu}}\mathrm{eV}$ Range with a Coaxial Dish Antenna},
  author = {Hoshino, Gabe and Knirck, Stefan and Awida, Mohamed H. and Cancelo, Gustavo I. and Corrodi, Simon and Di Federico, Martin and Knepper, Benjamin and Lapuente, Alex and Littmann, Mira and Miller, David W. and Mitchell, Donald V. and Rodriguez, Derrick and Ruschman, Mark K. and Salemi, Chiara P. and Sawtell, Matthew A. and Stefanazzi, Leandro and Sonnenschein, Andrew and Teafoe, Gary W. and Winter, Peter},
  collaboration = {GigaBREAD Collaboration},
  journal = {Phys. Rev. Lett.},
  volume = {134},
  issue = {17},
  pages = {171002},
  numpages = {9},
  year = {2025},
  month = {Apr},
  publisher = {American Physical Society},
  doi = {10.1103/PhysRevLett.134.171002},
  url = {https://link.aps.org/doi/10.1103/PhysRevLett.134.171002}
}

@article{Caputo2021,
  title = {Dark photon limits: A handbook},
  author = {Caputo, Andrea and Millar, Alexander J. and O'Hare, Ciaran A. J. and Vitagliano, Edoardo},
  journal = {Phys. Rev. D},
  volume = {104},
  issue = {9},
  pages = {095029},
  numpages = {35},
  year = {2021},
  month = {Nov},
  publisher = {American Physical Society},
  doi = {10.1103/PhysRevD.104.095029},
  url = {https://link.aps.org/doi/10.1103/PhysRevD.104.095029}
}

@article{GigaBREAD2024_PRL,
  title = {First Results from a Broadband Search for Dark Photon Dark Matter in the 44 to $52\text{ }\text{ }\mathrm{\ensuremath{\mu}}\mathrm{eV}$ Range with a Coaxial Dish Antenna},
  author = {Knirck, Stefan and Hoshino, Gabe and Awida, Mohamed H. and Cancelo, Gustavo I. and Di Federico, Martin and Knepper, Benjamin and Lapuente, Alex and Littmann, Mira and Miller, David W. and Mitchell, Donald V. and Rodriguez, Derrick and Ruschman, Mark K. and Sawtell, Matthew A. and Stefanazzi, Leandro and Sonnenschein, Andrew and Teafoe, Gary W. and Bowring, Daniel and Carosi, G. and Chou, Aaron and Chang, Clarence L. and Dona, Kristin and Khatiwada, Rakshya and Kurinsky, Noah A. and Liu, Jesse and Pena, Cristi\'an and Salemi, Chiara P. and Wang, Christina W. and Yu, Jialin},
  collaboration = {BREAD Collaboration},
  journal = {Phys. Rev. Lett.},
  volume = {132},
  issue = {13},
  pages = {131004},
  numpages = {6},
  year = {2024},
  month = {Mar},
  publisher = {American Physical Society},
  doi = {10.1103/PhysRevLett.132.131004},
  url = {https://link.aps.org/doi/10.1103/PhysRevLett.132.131004}
}

@article{QUALIPHIDE2023,
  title = {Wideband Direct Detection Constraints on Hidden Photon Dark Matter with the QUALIPHIDE Experiment},
  author = {Ramanathan, K. and Klimovich, N. and Basu Thakur, R. and Eom, B. H. and Leduc, H. G. and Shu, S. and Beyer, A. D. and Day, P. K.},
  journal = {Phys. Rev. Lett.},
  volume = {130},
  issue = {23},
  pages = {231001},
  numpages = {7},
  year = {2023},
  month = {Jun},
  publisher = {American Physical Society},
  doi = {10.1103/PhysRevLett.130.231001},
  url = {https://link.aps.org/doi/10.1103/PhysRevLett.130.231001}
}

@article{SHUKET2019,
  title = {Direct Searches for Hidden-Photon Dark Matter with the SHUKET Experiment},
  author = {Brun, Pierre and Chevalier, Laurent and Flouzat, Christophe},
  journal = {Phys. Rev. Lett.},
  volume = {122},
  issue = {20},
  pages = {201801},
  numpages = {5},
  year = {2019},
  month = {May},
  publisher = {American Physical Society},
  doi = {10.1103/PhysRevLett.122.201801},
  url = {https://link.aps.org/doi/10.1103/PhysRevLett.122.201801}
}

@article{Lewin1996_SHM,
  author = {Lewin, J. D. and Smith, P. F.},
  title = {Review of mathematics, numerical factors, and corrections for dark matter experiments based on elastic nuclear recoil},
  journal = {Astroparticle Physics},
  volume = {6},
  number = {1},
  pages = {87--112},
  year = {1996},
  doi = {10.1016/S0927-6505(96)00047-3}
}

@article{ADMX2021,
  title = {Search for Invisible Axion Dark Matter in the $3.3--4.2\text{ }\text{ }\ensuremath{\mu}\mathrm{eV}$ Mass Range},
  author = {Bartram, C. and Braine, T. and Burns, E. and Cervantes, R. and Crisosto, N. and Du, N. and Korandla, H. and Leum, G. and Mohapatra, P. and Nitta, T. and Rosenberg, L. J and Rybka, G. and Yang, J. and Clarke, John and Siddiqi, I. and Agrawal, A. and Dixit, A. V. and Awida, M. H. and Chou, A. S. and Hollister, M. and Knirck, S. and Sonnenschein, A. and Wester, W. and Gleason, J. R. and Hipp, A. T. and Jois, S. and Sikivie, P. and Sullivan, N. S. and Tanner, D. B. and Lentz, E. and Khatiwada, R. and Carosi, G. and Robertson, N. and Woollett, N. and Duffy, L. D. and Boutan, C. and Jones, M. and LaRoque, B. H. and Oblath, N. S. and Taubman, M. S. and Daw, E. J. and Perry, M. G. and Buckley, J. H. and Gaikwad, C. and Hoffman, J. and Murch, K. W. and Goryachev, M. and McAllister, B. T. and Quiskamp, A. and Thomson, C. and Tobar, M. E.},
  collaboration = {ADMX Collaboration},
  journal = {Phys. Rev. Lett.},
  volume = {127},
  issue = {26},
  pages = {261803},
  numpages = {6},
  year = {2021},
  month = {Dec},
  publisher = {American Physical Society},
  doi = {10.1103/PhysRevLett.127.261803},
  url = {https://link.aps.org/doi/10.1103/PhysRevLett.127.261803}
}

@article{HAYSTAC2023,
  title = {New results from HAYSTAC's phase II operation with a squeezed state receiver},
  author = {Jewell, M. J. and Leder, A. F. and Backes, K. M. and Bai, Xiran and van Bibber, K. and Brubaker, B. M. and Cahn, S. B. and Droster, A. and Esmat, Maryam H. and Ghosh, Sumita and Graham, Eleanor and Hilton, Gene C. and Jackson, H. and Laffan, Claire and Lamoreaux, S. K. and Lehnert, K. W. and Lewis, S. M. and Malnou, M. and Maruyama, R. H. and Palken, D. A. and Rapidis, N. M. and Ruddy, E. P. and Simanovskaia, M. and Singh, Sukhman and Speller, D. H. and Vale, Leila R. and Wang, H. and Zhu, Yuqi},
  collaboration = {HAYSTAC Collaboration},
  journal = {Phys. Rev. D},
  volume = {107},
  issue = {7},
  pages = {072007},
  numpages = {20},
  year = {2023},
  month = {Apr},
  publisher = {American Physical Society},
  doi = {10.1103/PhysRevD.107.072007},
  url = {https://link.aps.org/doi/10.1103/PhysRevD.107.072007}
}

@article{QUAX2023,
  title = {Search for galactic axions with a traveling wave parametric amplifier},
  author = {Di Vora, R. and Lombardi, A. and Ortolan, A. and Pengo, R. and Ruoso, G. and Braggio, C. and Carugno, G. and Taffarello, L. and Cappelli, G. and Crescini, N. and Esposito, M. and Planat, L. and Ranadive, A. and Roch, N. and Alesini, D. and Babusci, D. and D'Elia, A. and Di Gioacchino, D. and Gatti, C. and Ligi, C. and Maccarrone, G. and Rettaroli, A. and Tocci, S. and D'Agostino, D. and Gambardella, U. and Iannone, G. and Falferi, P.},
  collaboration = {QUAX Collaboration},
  journal = {Phys. Rev. D},
  volume = {108},
  issue = {6},
  pages = {062005},
  numpages = {13},
  year = {2023},
  month = {Sep},
  publisher = {American Physical Society},
  doi = {10.1103/PhysRevD.108.062005},
  url = {https://link.aps.org/doi/10.1103/PhysRevD.108.062005}
}

@article{DOSUE2023,
  title = {Search for dark photon dark matter in the mass range $41\ensuremath{-}74\text{ }\text{ }\mathrm{\ensuremath{\mu}}\mathrm{eV}$ using millimeter-wave receiver and radioshielding box},
  author = {Adachi, S. and Fujinaka, R. and Muto, Y. and Nakata, H. and Sueno, Y. and Sumida, T. and Suzuki, J. and Tajima, O. and Takeuchi, H. and Honda, S.},
  collaboration = {DOSUE-RR Collaboration},
  journal = {Phys. Rev. D},
  volume = {109},
  issue = {1},
  pages = {012008},
  numpages = {8},
  year = {2024},
  month = {Jan},
  publisher = {American Physical Society},
  doi = {10.1103/PhysRevD.109.012008},
  url = {https://link.aps.org/doi/10.1103/PhysRevD.109.012008}
}

@article{ORGAN2017,
   title={The ORGAN experiment: An axion haloscope above 15 GHz},
   volume={18},
   ISSN={2212-6864},
   url={http://dx.doi.org/10.1016/j.dark.2017.09.010},
   DOI={10.1016/j.dark.2017.09.010},
   journal={Physics of the Dark Universe},
   publisher={Elsevier BV},
   author={McAllister, Ben T. and Flower, Graeme and Ivanov, Eugene N. and Goryachev, Maxim and Bourhill, Jeremy and Tobar, Michael E.},
   year={2017},
   month=Dec, pages={67–72} }

@article{Cowan2011,
   title={Asymptotic formulae for likelihood-based tests of new physics},
   volume={71},
   ISSN={1434-6052},
   url={http://dx.doi.org/10.1140/epjc/s10052-011-1554-0},
   DOI={10.1140/epjc/s10052-011-1554-0},
   number={2},
   journal={The European Physical Journal C},
   publisher={Springer Science and Business Media LLC},
   author={Cowan, Glen and Cranmer, Kyle and Gross, Eilam and Vitells, Ofer},
   year={2011},
   month=Feb }

@article{PDG2024,
  title = {Review of Particle Physics},
  author = {Navas, S. and Amsler, C. and Gutsche, T. and Hanhart, C. and Hern\'andez-Rey, J. J. and Louren\ifmmode \mbox{\c{c}}\else \c{c}\fi{}o, C. and Masoni, A. and Mikhasenko, M. and Mitchell, R. E. and Patrignani, C. and Schwanda, C. and Spanier, S. and Venanzoni, G. and Yuan, C. Z. and Agashe, K. and Aielli, G. and Allanach, B. C. and Alvarez-Mu\~niz, J. and Antonelli, M. and Aschenauer, E. C. and Asner, D. M. and Assamagan, K. and Baer, H. and Banerjee, Sw. and Barnett, R. M. and Baudis, L. and Bauer, C. W. and Beatty, J. J. and Beringer, J. and Bettini, A. and Biebel, O. and Black, K. M. and Blucher, E. and Bonventre, R. and Briere, R. A. and Buckley, A. and Burkert, V. D. and Bychkov, M. A. and Cahn, R. N. and Cao, Z. and Carena, M. and Casarosa, G. and Ceccucci, A. and Cerri, A. and Chivukula, R. S. and Cowan, G. and Cranmer, K. and Crede, V. and Cremonesi, O. and D'Ambrosio, G. and Damour, T. and de Florian, D. and de Gouv\^ea, A. and DeGrand, T. and Demers, S. and Demiragli, Z. and Dobrescu, B. A. and D'Onofrio, M. and Doser, M. and Dreiner, H. K. and Eerola, P. and Egede, U. and Eidelman, S. and El-Khadra, A. X. and Ellis, J. and Eno, S. C. and Erler, J. and Ezhela, V. V. and Fava, A. and Fetscher, W. and Fields, B. D. and Freitas, A. and Gallagher, H. and Gershon, T. and Gershtein, Y. and Gherghetta, T. and Gonzalez-Garcia, M. C. and Goodman, M. and Grab, C. and Gritsan, A. V. and Grojean, C. and Groom, D. E. and Gr\"unewald, M. and Gurtu, A. and Haber, H. E. and Hamel, M. and Hashimoto, S. and Hayato, Y. and Hebecker, A. and Heinemeyer, S. and Hikasa, K. and Hisano, J. and H\"ocker, A. and Holder, J. and Hsu, L. and Huston, J. and Hyodo, T. and Ianni, Al. and Kado, M. and Karliner, M. and Katz, U. F. and Kenzie, M. and Khoze, V. A. and Klein, S. R. and Krauss, F. and Kreps, M. and Kri\ifmmode \check{z}\else \v{z}\fi{}an, P. and Krusche, B. and Kwon, Y. and Lahav, O. and Lellouch, L. P. and Lesgourgues, J. and Liddle, A. R. and Ligeti, Z. and Lin, C.-J. and Lippmann, C. and Liss, T. M. and Lister, A. and Littenberg, L. and Lugovsky, K. S. and Lugovsky, S. B. and Lusiani, A. and Makida, Y. and Maltoni, F. and Manohar, A. V. and Marciano, W. J. and Matthews, J. and Mei\ss{}ner, U.-G. and Melzer-Pellmann, I.-A. and Mertsch, P. and Miller, D. J. and Milstead, D. and M\"onig, K. and Molaro, P. and Moortgat, F. and Moskovic, M. and Nagata, N. and Nakamura, K. and Narain, M. and Nason, P. and Nelles, A. and Neubert, M. and Nir, Y. and O'Connell, H. B. and O'Hare, C. A. J. and Olive, K. A. and Peacock, J. A. and Pianori, E. and Pich, A. and Piepke, A. and Pietropaolo, F. and Pomarol, A. and Pordes, S. and Profumo, S. and Quadt, A. and Rabbertz, K. and Rademacker, J. and Raffelt, G. and Ramsey-Musolf, M. and Richardson, P. and Ringwald, A. and Robinson, D. J. and Roesler, S. and Rolli, S. and Romaniouk, A. and Rosenberg, L. J and Rosner, J. L. and Rybka, G. and Ryskin, M. G. and Ryutin, R. A. and Safdi, B. and Sakai, Y. and Sarkar, S. and Sauli, F. and Schneider, O. and Sch\"onert, S. and Scholberg, K. and Schwartz, A. J. and Schwiening, J. and Scott, D. and Sefkow, F. and Seljak, U. and Sharma, V. and Sharpe, S. R. and Shiltsev, V. and Signorelli, G. and Silari, M. and Simon, F. and Sj\"ostrand, T. and Skands, P. and Skwarnicki, T. and Smoot, G. F. and Soffer, A. and Sozzi, M. S. and Spiering, C. and Stahl, A. and Sumino, Y. and Takahashi, F. and Tanabashi, M. and Tanaka, J. and Ta\ifmmode \check{s}\else \v{s}\fi{}evsk\'y, M. and Terao, K. and Terashi, K. and Terning, J. and Thoma, U. and Thorne, R. S. and Tiator, L. and Titov, M. and Tovey, D. R. and Trabelsi, K. and Urquijo, P. and Valencia, G. and Van de Water, R. and Varelas, N. and Verde, L. and Vivarelli, I. and Vogel, P. and Vogelsang, W. and Vorobyev, V. and Wakely, S. P. and Walkowiak, W. and Walter, C. W. and Wands, D. and Weinberg, D. H. and Weinberg, E. J. and Wermes, N. and White, M. and Wiencke, L. R. and Willocq, S. and Woody, C. L. and Workman, R. L. and Yao, W.-M. and Yokoyama, M. and Yoshida, R. and Zanderighi, G. and Zeller, G. P. and Zhu, R.-Y. and Zhu, S.-L. and Zimmermann, F. and Zyla, P. A. and Anderson, J. and Kramer, M. and Schaffner, P. and Zheng, W.},
  collaboration = {Particle Data Group Collaboration},
  journal = {Phys. Rev. D},
  volume = {110},
  issue = {3},
  pages = {030001},
  numpages = {5},
  year = {2024},
  month = {Aug},
  publisher = {American Physical Society},
  doi = {10.1103/PhysRevD.110.030001},
  url = {https://link.aps.org/doi/10.1103/PhysRevD.110.030001}
}

@book{Cowan1998,
  author = {Cowan, Glen},
  title = {Statistical Data Analysis},
  publisher = {Oxford University Press},
  year = {1998},
  address = {Oxford},
  isbn = {978-0198501558}
}

@book{Sivia2006,
  author = {Sivia, D. S. and Skilling, J.},
  title = {Data Analysis: A Bayesian Tutorial},
  publisher = {Oxford University Press},
  year = {2006},
  edition = {2nd},
  address = {Oxford},
  isbn = {978-0198568322}
}

@article{Read2014,
   title={The local dark matter density},
   volume={41},
   ISSN={1361-6471},
   url={http://dx.doi.org/10.1088/0954-3899/41/6/063101},
   DOI={10.1088/0954-3899/41/6/063101},
   number={6},
   journal={Journal of Physics G: Nuclear and Particle Physics},
   publisher={IOP Publishing},
   author={Read, J I},
   year={2014},
   month={May}, pages={063101} }

@misc{AxionLimits,
  author = {O'Hare, C. A. J.},
  title = {{AxionLimits}},
  year = {2020},
  howpublished = {\url{https://github.com/cajohare/AxionLimits}},
  note = {github.com/cajohare/AxionLimits}
}

@ARTICLE{RubinFord1970,
       author = {{Rubin}, Vera C. and {Ford}, Jr., W. Kent},
        title = "{Rotation of the Andromeda Nebula from a Spectroscopic Survey of Emission Regions}",
      journal = {\apj},
         year = 1970,
        month = feb,
       volume = {159},
        pages = {379},
          doi = {10.1086/150317},
       adsurl = {https://ui.adsabs.harvard.edu/abs/1970ApJ...159..379R}
}

@ARTICLE{Tyson1998,
       author = {{Tyson}, J. Anthony and {Kochanski}, Greg P. and {Dell'Antonio}, Ian P.},
        title = "{Detailed Mass Map of CL 0024+1654 from Strong Lensing}",
      journal = {Astrophys. J. Lett.},
         year = 1998,
        month = may,
       volume = {498},
       number = {2},
        pages = {L107-L110},
          doi = {10.1086/311314},
archivePrefix = {arXiv},
       eprint = {astro-ph/9801193},
 primaryClass = {astro-ph},
       adsurl = {https://ui.adsabs.harvard.edu/abs/1998ApJ...498L.107T}
}

@article{Tegmark2004,
  title = {Cosmological parameters from SDSS and WMAP},
  author = {Tegmark, Max and Strauss, Michael A. and Blanton, Michael R. and Abazajian, Kevork and Dodelson, Scott and Sandvik, Havard and Wang, Xiaomin and Weinberg, David H. and Zehavi, Idit and Bahcall, Neta A. and Hoyle, Fiona and Schlegel, David and Scoccimarro, Roman and Vogeley, Michael S. and Berlind, Andreas and Budavari, Tam{\'a}s and Connolly, Andrew and Eisenstein, Daniel J. and Finkbeiner, Douglas and Frieman, Joshua A. and Gunn, James E. and Hui, Lam and Jain, Bhuvnesh and Johnston, David and Kent, Stephen and Lin, Huan and Nakajima, Reiko and Nichol, Robert C. and Ostriker, Jeremiah P. and Pope, Adrian and Scranton, Ryan and Seljak, Uro{\ifmmode \check{s}\else \v{s}\fi{}} and Sheth, Ravi K. and Stebbins, Albert and Szalay, Alexander S. and Szapudi, Istv{\'a}n and Xu, Yongzhong and Annis, James and Brinkmann, J. and Burles, Scott and Castander, Francisco J. and Csabai, Istvan and Loveday, Jon and Doi, Mamoru and Fukugita, Masataka and Gillespie, Bruce and Hennessy, Greg and Hogg, David W. and Ivezi{\ifmmode \acute{c}\else \'{c}\fi{}}, {\ifmmode \check{Z}\else \v{Z}\fi{}}eljko and Knapp, Gillian R. and Lamb, Don Q. and Lee, Brian C. and Lupton, Robert H. and McKay, Timothy A. and Kunszt, Peter and Munn, Jeffrey A. and O'Connell, Liam and Peoples, John and Pier, Jeffrey R. and Richmond, Michael and Rockosi, Constance and Schneider, Donald P. and Stoughton, Christopher and Tucker, Douglas L. and Vanden Berk, Daniel E. and Yanny, Brian and York, Donald G.},
  journal = {Phys. Rev. D},
  volume = {69},
  issue = {10},
  pages = {103501},
  numpages = {26},
  year = {2004},
  month = {May},
  publisher = {American Physical Society},
  doi = {10.1103/PhysRevD.69.103501},
  url = {https://link.aps.org/doi/10.1103/PhysRevD.69.103501}
}

@article{HAYSTAC_procedure,
  title = {HAYSTAC axion search analysis procedure},
  author = {Brubaker, B. M. and Zhong, L. and Lamoreaux, S. K. and Lehnert, K. W. and van Bibber, K. A.},
  journal = {Phys. Rev. D},
  volume = {96},
  issue = {12},
  pages = {123008},
  numpages = {34},
  year = {2017},
  month = {Dec},
  publisher = {American Physical Society},
  doi = {10.1103/PhysRevD.96.123008},
  url = {https://link.aps.org/doi/10.1103/PhysRevD.96.123008}
}

@article{HAYSTAC2026,
  title = {Search for dark photons between $16.96--19.52\text{ }\text{ }\mathrm{\ensuremath{\mu}}\mathrm{eV}$ with the HAYSTAC experiment},
  author = {Bai, Xiran and Droster, A. and Echevers, J. and Esmat, Maryam H. and Ghosh, Sumita and Graham, Eleanor and Jackson, H. and Jois, S. and Jewell, M. J. and Laffan, Claire and Leder, A. F. and Lehnert, K. W. and Lewis, S. M. and Maruyama, R. H. and Rapidis, N. M. and Ruddy, E. P. and Silva-Feaver, M. and Simanovskaia, M. and Singh, Sukhman and Speller, D. H. and van Bibber, K. and Wang, Y. and Zacarias, Sabrina and Zhu, Yuqi},
  collaboration = {HAYSTAC Collaboration},
  journal = {Phys. Rev. D},
  volume = {113},
  issue = {1},
  pages = {012010},
  numpages = {6},
  year = {2026},
  month = {Jan},
  publisher = {American Physical Society},
  doi = {10.1103/3txs-dz63},
  url = {https://link.aps.org/doi/10.1103/3txs-dz63}
}

@article{DFSZ2023,
  title = {Axion Dark Matter Search around $4.55\text{ }\text{ }\mathrm{\ensuremath{\mu}}\mathrm{eV}$ with Dine-Fischler-Srednicki-Zhitnitskii Sensitivity},
  author = {Yi, Andrew K. and Ahn, Saebyeok and Kutlu, \ifmmode \mbox{\c{C}}\else \c{C}\fi{}a\ifmmode \breve{g}\else \u{g}\fi{}lar and Kim, JinMyeong and Ko, Byeong Rok and Ivanov, Boris I. and Byun, HeeSu and van Loo, Arjan F. and Park, SeongTae and Jeong, Junu and Kwon, Ohjoon and Nakamura, Yasunobu and Uchaikin, Sergey V. and Choi, Jihoon and Lee, Soohyung and Lee, MyeongJae and Shin, Yun Chang and Kim, Jinsu and Lee, Doyu and Ahn, Danho and Bae, SungJae and Lee, Jiwon and Kim, Younggeun and Gkika, Violeta and Lee, Ki Woong and Oh, Seonjeong and Seong, Taehyeon and Kim, DongMin and Chung, Woohyun and Matlashov, Andrei and Youn, SungWoo and Semertzidis, Yannis K.},
  journal = {Phys. Rev. Lett.},
  volume = {130},
  issue = {7},
  pages = {071002},
  numpages = {7},
  year = {2023},
  month = {Feb},
  publisher = {American Physical Society},
  doi = {10.1103/PhysRevLett.130.071002},
  url = {https://link.aps.org/doi/10.1103/PhysRevLett.130.071002}
}

@article{Suzuki_2015,
   title={Experimental search for hidden photon CDM in the eV mass range with a dish antenna},
   volume={2015},
   ISSN={1475-7516},
   url={http://dx.doi.org/10.1088/1475-7516/2015/09/042},
   DOI={10.1088/1475-7516/2015/09/042},
   number={09},
   journal={Journal of Cosmology and Astroparticle Physics},
   publisher={IOP Publishing},
   author={Suzuki, J. and Horie, T. and Inoue, Y. and Minowa, M.},
   year={2015},
   month=sep , pages={042–042} }

@article{Tomita_2020,
   title={Search for hidden-photon cold dark matter using a K-band cryogenic receiver},
   volume={2020},
   ISSN={1475-7516},
   url={http://dx.doi.org/10.1088/1475-7516/2020/09/012},
   DOI={10.1088/1475-7516/2020/09/012},
   number={09},
   journal={Journal of Cosmology and Astroparticle Physics},
   publisher={IOP Publishing},
   author={Tomita, N. and Oguri, S. and Inoue, Y. and Minowa, M. and Nagasaki, T. and Suzuki, J. and Tajima, O.},
   year={2020},
   month= sep, pages={012–012} }

@article{BRASS-p2023,
doi = {10.1088/1475-7516/2023/08/077},
url = {https://doi.org/10.1088/1475-7516/2023/08/077},
year = {2023},
month = {aug},
publisher = {IOP Publishing},
volume = {2023},
number = {08},
pages = {077},
author = {Bajjali, Fayez and Dornbusch, Sven and Ekmedžić, Marko and Horns, Dieter and Kasemann, Christoph and Lobanov, Andrei and Mkrtchyan, Artak and Nguyen, Le Hoang and Tluczykont, Martin and Tuccari, Gino and Ulrichs, Johannes and Wieching, Gundolf and Zensus, Anton},
title = {First results from BRASS-p broadband searches for hidden photon dark matter},
journal = {Journal of Cosmology and Astroparticle Physics}
}

@article{CMWR2023,
  title = {Search for Dark Photon Dark Matter in the Mass Range $74--110\text{ }\text{ }\mathrm{\ensuremath{\mu}}\mathrm{eV}$ with a Cryogenic Millimeter-Wave Receiver},
  author = {Kotaka, S. and Adachi, S. and Fujinaka, R. and Honda, S. and Nakata, H. and Seino, Y. and Sueno, Y. and Sumida, T. and Suzuki, J. and Tajima, O. and Takeichi, S.},
  collaboration = {DOSUE-RR Collaboration},
  journal = {Phys. Rev. Lett.},
  volume = {130},
  issue = {7},
  pages = {071805},
  numpages = {5},
  year = {2023},
  month = {Feb},
  publisher = {American Physical Society},
  doi = {10.1103/PhysRevLett.130.071805},
  url = {https://link.aps.org/doi/10.1103/PhysRevLett.130.071805}
}

@article{CAPP2024,
  title = {Extensive Search for Axion Dark Matter over 1 GHz with CAPP'S Main Axion Experiment},
  author = {Ahn, Saebyeok and Kim, JinMyeong and Ivanov, Boris I. and Kwon, Ohjoon and Byun, HeeSu and van Loo, Arjan F. and Park, SeongTae and Jeong, Junu and Lee, Soohyung and Kim, Jinsu and Kutlu, \ifmmode \mbox{\c{C}}\else \c{C}\fi{}a\ifmmode \breve{g}\else \u{g}\fi{}lar and Yi, Andrew K. and Nakamura, Yasunobu and Oh, Seonjeong and Ahn, Danho and Bae, SungJae and Choi, Hyoungsoon and Choi, Jihoon and Chong, Yonuk and Chung, Woohyun and Gkika, Violeta and Kim, Jihn E. and Kim, Younggeun and Ko, Byeong Rok and Miceli, Lino and Lee, Doyu and Lee, Jiwon and Lee, Ki Woong and Lee, MyeongJae and Matlashov, Andrei and Parashar, Pallavi and Seong, Taehyeon and Shin, Yun Chang and Uchaikin, Sergey V. and Youn, SungWoo and Semertzidis, Yannis K.},
  journal = {Phys. Rev. X},
  volume = {14},
  issue = {3},
  pages = {031023},
  numpages = {32},
  year = {2024},
  month = {Aug},
  publisher = {American Physical Society},
  doi = {10.1103/PhysRevX.14.031023},
  url = {https://link.aps.org/doi/10.1103/PhysRevX.14.031023}
}

@article{ORGAN_2023,
   title={Limits on Dark Photons, Scalars, and Axion‐Electromagnetodynamics with the ORGAN Experiment},
   volume={536},
   ISSN={1521-3889},
   url={http://dx.doi.org/10.1002/andp.202200622},
   DOI={10.1002/andp.202200622},
   number={1},
   journal={Annalen der Physik},
   publisher={Wiley},
   author={McAllister, Ben T. and Quiskamp, Aaron and O’Hare, Ciaran A. J. and Altin, Paul and Ivanov, Eugene N. and Goryachev, Maxim and Tobar, Michael E.},
   year={2023},
   month={June} }

@article{PhysRevD.110.043022,
  title = {Deepest sensitivity to wavelike dark photon dark matter with superconducting radio frequency cavities},
  author = {Cervantes, R. and Aumentado, J. and Braggio, C. and Giaccone, B. and Frolov, D. and Grassellino, A. and Harnik, R. and Lecocq, F. and Melnychuk, O. and Pilipenko, R. and Posen, S. and Romanenko, A.},
  journal = {Phys. Rev. D},
  volume = {110},
  issue = {4},
  pages = {043022},
  numpages = {10},
  year = {2024},
  month = {Aug},
  publisher = {American Physical Society},
  doi = {10.1103/PhysRevD.110.043022},
  url = {https://link.aps.org/doi/10.1103/PhysRevD.110.043022}
}
\end{document}